\documentclass[11pt,a4paper]{article}
\usepackage{jheppub}
\usepackage{amssymb,amsmath,amsfonts,mathrsfs}
\usepackage{graphicx}
\usepackage[dvipsnames]{xcolor}
\usepackage{verbatim}
\usepackage{epsfig}
\usepackage{color}
\usepackage{multirow}
\usepackage{comment}
\usepackage{datetime}
\usepackage{slashed}
\usepackage{ulem}
\usepackage{framed}
\usepackage{longtable}
\usepackage[mathscr]{euscript}
\usepackage{cancel}
\usepackage{tensor}
 \usepackage{mathtools}
  \usepackage{caption}
  \usepackage{subcaption}
 \usepackage[multiple]{footmisc}
 \usepackage{pgfplots}
 \usepackage{natbib,bibentry}
\usepackage[applemac]{inputenc}
\usepackage{tikz}
\usetikzlibrary{decorations.markings}

\newcommand{\nn}{\nonumber}

\def\be{\begin{equation}}
\def\ee{\end{equation}}
\def\bea{\begin{align}}
\def\eea{\end{align}}

\def\a{\alpha}		\def\b{\beta}		\def\g{\gamma}		\def\d{\delta}
\def\e{\varepsilon}	\def\z{\zeta}					\def\q{\theta}
						
\def\n{\nu}									\def\r{\rho}	
\def\s{\sigma}		\def\t{\tau}			\def\f{\phi}			
\def\c{\chi}

\title{Turning rotating D-branes and BHs inside out their photon-halo}
\author{Massimo Bianchi,}
\author{Giorgio Di Russo}

\affiliation{Dipartimento di Fisica,  Universit\`a di Roma ``Tor Vergata"  \& Sezione INFN Roma2, Via della ricerca scientifica 1, 00133, Roma, Italy}

\abstract{
We extend our investigation on Couch-Torrence conformal inversions of BHs and D-branes in various directions. We analyse asymptotically flat rotating charged BHs in $D=4$, in particular extremal rotating BHs in STU supergavity, and find invariance for special choices of the charges. Due to the dependence of the critical radii on the impact parameter(s), the relation between the scattering angle for geodesics outside the photon-halo and the in-spiralling angle for geodesics inside the photon-halo is modified by the inclusion of a boundary term. We also consider rotating BHs in $D=5$ and rotating D3-branes and find invariance under generalised Couch-Torrence inversions for special choices of the angular momenta. Alas we don't find any similar symmetry for smooth horizonless geometries. Moreover, relying on the surprising connection between classical BH perturbation theory and quantum Seiberg-Witten curves for ${\cal N}=2$ SYM theories, we study scalar wave equations in these backgrounds and identify the near super-radiant modes produced in near-extremal BH mergers. Finally, we study scalar fluctuations around Kerr-Newman BHs in AdS$_4$ and find stringent conditions for generalised Couch-Torrence symmetry that are relaxed in the extremal case or by allowing a rescaling of the wave function.}

\begin{document}
\maketitle
\flushbottom 
\section{Introduction}
Many compact gravitating objects are surrounded by a ring or a halo of light formed by the quasi-critical geodesics of photons surfing the potential barrier, called the photon-sphere or photon-halo, separating the horizon from the asymptotically flat region\cite{Cardoso:2017cqb, Cardoso:2019rvt}. Perturbations of these barriers reflect into the quasi-normal modes (QNMs) that dominate the gravitational wave (GW) signal during the prompt ring-down phase that follows the merger of two black-holes (BHs), for instance \cite{Barack:2018yly, Barausse:2020rsu}. 

In \cite{Bianchi:2021yqs} we have shown that D3-branes and their bound-states in lower dimensions admit a symmetry under conformal inversions that generalise the Couch-Torrence (CT) transformations known to leave the metric of extremal Reissner-Nordstr\"om BHs invariant up to a Weyl rescaling \cite{CouchTorr}. CT transformations are known to exchange the event horizon with null infinity \cite{CouchTorr,Aretakis:2011ha, Aretakis:2011hc, Aretakis:2012ei, Godazgar:2017igz, Godazgar:2019dkh, Cvetic:2020kwf, Cvetic:2021lss}. Quite remarkably, we have found that the fixed loci of the (generalised) CT transformations are precisely the photon-spheres \cite{Bianchi:2021yqs}, thus opening new paths to explore physical implications for QNMs and other observables such as deflection angles and time delays, coded in the so-called radial action, and ultimately on a bulk-to-boundary relation proposed in \cite{Kalin:2019rwq, Kalin:2019inp}. 

In the present investigation we will extend our analysis to the case of rotating compact objects. In the extremal case (zero temperature), Kerr and Kerr-Newman (KN) BHs \cite{CouchTorr} as well as their cousins in STU supergravity (for special choices of the charges) \cite{Cvetic:2020kwf, Cvetic:2021lss} are known to admit a remnant of the CT inversion symmetry. Although the metric is not conformally invariant, the radial equation, that results from the separability of the dynamics, is invariant under transformations that depend on the angular momentum of the perturbation (or the impact parameter of probe). Even more remarkably we will show that the fixed loci of the conformal inversions are the photon-halos that form the `asymmetric' light-ring structures \cite{Himwich:2020msm} familiar from the images produced by the EHT collaboration \cite{EventHorizonTelescope:2019dse}. We will confirm that only special choices of the charges allow for the symmetry in its elementary form. Otherwise a Freudenthal duality transformation of the charges would be required \cite{Borsten:2018djw, Borsten:2019xas}.  We then study the implications for other physical observables and find that the radial action is formally invariant but, even in the equatorial plane $\theta=\pi/2$, the scattering angle $\Delta\phi_{scatt}$ for a probe impinging from outside the photon-halo cannot be simply related to the in spiraling  angle $\Delta\phi_{fall}$ of a probe with the same energy $E$ and angular momentum $J$ falling into the horizon. We will explain this discrepancy in terms of the dependence of the extrema of integration on the impact parameter $b=J/E$ and discuss the issue for generic non-planar motion. We also study scalar fluctuations in the (near extremal) STURBHs (rotating BH in STU supergravity) and identify the near super-radiant modes that are long-lived in that their frequencies have a very small imaginary parts. 

We then address the fate of the CT inversions for rotating D3-branes \cite{Russo:1998by} and conclude that generalised CT transformations are symmetries of massless geodesics, at least for special choices of the rotation parameters. Extremal rotating objects do enjoy CT invariance beyond $d=4$ in the extremal (not necessarily BPS) case. We illustrate our conclusions in the case of rotating BHs in $d=5$ of the BMPV supersymmetric BPS kind \cite{Breckenridge:1996is} as well as their non-supersymmetric cousins of the CCLP family \cite{Chong:2004na, Chong:2005hr}. However their smooth horizonless uplifts to $d=6$ known as JMaRT solutions \cite{Jejjala:2005yu} or GMS \cite{Giusto:2004id, Giusto:2004ip} `fake' fuzz-balls (in the BPS case), lacking a horizon, fail to admit any symmetry under conformal inversions, even though they are endowed with a photon-halo in most cases. A similar story applies to circular fuzz-balls that represent a class of micro-state geometries of two charge systems like D3-D3' (or D1-D5 after T-duality). Yet some embryonic form of invariance property is enjoyed by special classes of geodesics in a restricted subspace (e.g. $\theta=0$ plane, orthogonal to the circle).

Finally we consider KN BHs in AdS$_4$ and determine the conditions for CT invariance exploiting the recently established connection between BH perturbation theory and quantum Seiberg-Witten (SW) curves \cite{Aminov:2020yma, Bianchi:2021xpr, Bonelli:2021uvf, Bianchi:2021mft, Bonelli:2022ten}.

The plan of the paper is as follows. We start by reviewing the prototypical rotating case of the KN BH in Section \ref{KerrNew}. We focus on null geodesics as in \cite{Bianchi:2021yqs}  rather than on massless scalar wave perturbations as in \cite{CouchTorr, Cvetic:2020kwf, Cvetic:2021lss}. Some details on the critical parameters are given in Appendix \ref{appA}. 
%As mentioned above, we identify the fixed loci of generalised CT inversions with the photon-halos formed by the set of photon-rings at varying impact parameters $b={K\over E}$, $\zeta ={J\over E}{-}a$ for both co-rotating $Ja>0$ and counter-rotating $Ja<0$ geodesics. The radial action as well as the deflection angle and Shapiro time delay are shown to be `formally' invariant, up to subtleties that we will spell out.  

In Section \ref{eSTURBH} we then pass to consider eSTURBHs and show that the condition on the (four electric) charges for the perturbations to be invariant under GCT is to the one found in in \cite{Bianchi:2021yqs} for non-rotating BHs. 
The case $Q_1=Q_2=Q_3=Q_4=Q$ is shown to coincide with KN BHs. The details are presented in Appendix \ref{appB}.

Scalar wave fluctuations of near-eSTURBHs are studied in Section \ref{STUBHwave} and the frequency of the near SR modes are determined whose expressions take a very simple form when the conditions for CT invariance are met.  
  
Rotating BHs of the CCLP kind in $D=5$ and rotating D3-branes are discussed in Section \ref{RotD3}, while circular D3-D3' (or D1/D5) fuzz-balls are discussed in Section \ref{RotD3D3fuzz} together with JMaRT and GMs. 

The study of KN BHs in AdS and their connection with Heun Equations (HE) and quantum SW curves is the subject of Section \ref{AdSKNsec}.

Section \ref{concl} contains a summary of our results, the conclusions we draw from them and our outlook.

\section{Kerr-Newman BH}
\label{KerrNew}

Let us start with analysing CT inversions in the simplest cases of charged rotating BHs in $d=4$, namely KN BH. Rather than studying massless neutral scalar wave perturbations of the geometry as in \cite{CouchTorr, Cvetic:2020kwf, Cvetic:2021lss}, in this Section we will concentrate our attention on null geodesics as in \cite{Bianchi:2021yqs}. The first important result that we find is the identification of the fixed loci of generalised CT inversions with the photon-halos, formed by the collection of photon-rings at varying impact parameters. Indeed due to the preferred axis of the geometry, represented by the angular momentum $\vec{J}_{BH} = M \vec{a}$, one should introduce two independent impact parameters: $b={K/E}$, where $K$ is Carter's constant of separation \cite{ChandraBH}, related to the total angular momentum of the probe, and $b_J=J/E$, where $J$ is the projection of the angular momentum of the probe along $\vec{J}_{BH}$. Due to frame-dragging one should distinguish the two cases of co-rotating $Ja>0$ and counter-rotating $Ja<0$ geodesics. After the simplified analysis of equatorial motion ($\theta = \pi/2$, $J=\pm K$), that allows to `easily' show that the radial action as well as the deflection angle and Shapiro time delay are `formally' invariant, we tackle the subtleties involved with the dependence of the extrema of integration on the impact parameter and eventually address the general case of non-planar geodesics. Thanks to separability, the dynamics is encoded in an `angular' action $S_\theta$ in addition to the radial action $S_r$.    

The metric of KN space-time in Boyer-Lindquist coordinates and in natural units of $G_N=1$, $c=1$ reads \cite{ChandraBH}
\be
ds^2=-{\Delta_r \over \rho^2}\left(dt-a \sin^2\q d\phi\right)^2+{\sin^2\q\over\rho^2}\left(a dt-(a^2+r^2)d\phi\right)^2+{\rho^2 dr^2\over\Delta_r}+\rho^2 d\q^2
\ee
where
\be
\Delta_r=r^2+a^2-2 M r+Q^2,\quad\quad\rho^2=r^2+a^2\cos^2\q.
\ee
The horizons are the zeros of $\Delta_r$
\be
r_H^{\pm}=M\pm\sqrt{M^2-Q^2-a^2}.
\ee
while the singularity or rather `ringularity', since it is a `ring' in the equatorial  plane, is located at $\rho=0$, ie $r=0$, $\theta=\pi/2$. The external horizon is surrounded by an ergo-region, delimited by the ergo-sphere, wherein the time-like Killing vector $\partial_t$ becomes space-like thus leading to interesting processes like Penrose mechanism \cite{Penrose:1971uk} and super-radiance \cite{Brito:2015oca}, to which will come back later on for STURBH (rotating BHs in STU supergravity) \cite{Cvetic:1996xz}. 

When the extremality condition
\be
M^2=Q^2+a^2.
\ee
is met, $r_+=r_-=r_H = M = \sqrt{Q^2+a^2}$. For later use it is convenient to  
perform the following change of radial coordinate:
\be
\hat{r}=r-\sqrt{Q^2+a^2},
\ee
that maps the horizon into $\hat{r}_H=0$. In this coordinate the null Hamiltonian reads
\be
\mathcal{H}={1\over 2\hat{\rho}^2}\left\{P_{\hat{r}}^2\hat{r}^2-{1\over \hat{r}^2}{\left[E\left(a^2+\left(\hat{r}+\sqrt{a^2+Q^2}\right)^2\right)-aJ\right]^2}\right\}+{1\over 2\hat{\rho}^2}\left\{P_{\q}^2+{\left(a E \sin^2\q-J\right)^2\over \sin^2\q}\right\}
\ee
where $E={-}P_t$ and $P_{\phi}=J$ are conserved quantities.
% and 
%\be
%\hat{\rho}^2=(\hat{r}+\sqrt{a^2+Q^2})^2+a^2\cos^2\q .
%\ee
Following Carter \cite{ChandraBH}, the null condition $\mathcal{H}=0$ can be solved by introducing a `separation' constant $K$, representing the total angular momentum (including frame dragging) such that
\be\label{sep}
K^2=P_\q^2+{(a E \sin^2\q-J)^2\over \sin^2\q}=-P_{\hat{r}}^2\hat{r}^2+{1\over \hat{r}^2}{\left[E\left(a^2+\left(\hat{r}+\sqrt{a^2+Q^2}\right)^2\right)-aJ\right]^2}.
\ee
Defining  the impact parameters  
\be
\z={J\over E}-a\quad,\quad b={K\over E}
\ee
equations \eqref{sep} can be written in the form
\begin{align}\label{ptheta}
&P_{\hat{r}}^2 ={E^2\over \hat{r}^4  } \mathcal{R}(\hat{r})\quad {\rm with} \quad \mathcal{R}(\hat{r})=\left[\left(\hat{r}+\sqrt{a^2+Q^2}\right)^2-a\z\right]^2-b^2\hat{r}^2\\\nn
&P_{\q}^2  ={E^2\over \sin^2\q} \Theta(\cos\q) \quad {\rm with} \quad \Theta(\cos\q)= b^2\sin^2\q-\left(\z+a\cos^2\q\right)^2
\end{align}
with $\mathcal{R}$ and $\Theta$ quartic polynomials of $\hat{r}$ and $\chi=\cos\q$, respectively. 

\subsection{Geodesics in the equatorial plane}
Planar motion in the equatorial plane $\q=\pi/2$ is allowed  for $P_\theta = \rho^2 \dot\theta = 0$ i.e. for $b^2=\zeta^2$. In these cases \eqref{sep} reduces to\footnote{For simplicity here we indicate by $b=J/E=\pm K/E$ the relevant impact parameter that has a `sign', in that $ab>0$ represents co-rotating geodesics and $ab<0$ represents counter-rotating geodesics.}:
\be
\mathcal{R}(\hat{r}) = {P^2_{\hat{r}}\hat{r}^4\over E^2}=\prod_{i=1}^4(\hat{r}-\hat{r}_i)
\ee
where the roots are:
\be
\hat{r}_1= \hat{r}^{[+]}_{+}, \quad \hat{r}_2= \hat{r}^{[-]}_{+}, \quad 
\hat{r}_3= \hat{r}^{[+]}_{-}, \quad \hat{r}_4= \hat{r}^{[-]}_{-} 
\ee
with 
\be
\hat{r}^{[\pm]}_{\pm}={1\over 2}\left[\pm (a-b)-2\sqrt{a^2+Q^2}[\pm]\sqrt{-3a^2+2ab+b^2\pm 4\sqrt{a^2+Q^2}(b-a)}\right]
\ee
%nonumber\\
%\hat{r}_2&={1\over 2}\left[a-b-2\sqrt{a^2+Q^2}-\sqrt{-3a^2+2ab+b^2-4\sqrt{a^2+Q^2}(a-b)}\right]\nonumber\\
%\hat{r}_3&={1\over 2}\left[-a+b-2\sqrt{a^2+Q^2}+\sqrt{-3a^2+2ab+b^2+4\sqrt{a^2+Q^2}(a-b)}\right]\nonumber\\
%\hat{r}_4&={1\over 2}\left[-a+b-2\sqrt{a^2+Q^2}-\sqrt{-3a^2+2ab+b^2+4\sqrt{a^2+Q^2}(a-b)}\right].
%\end{align}
In the extremal case under consideration, the roots satisfy 
\be\label{r1r2betasquare}
\hat{r}_1\hat{r}_2=\hat{r}_3\hat{r}_4=2a^2-a b+Q^2 = \hat{r}_c^2.
\ee
that determine the critical radius as a function of the BH parameters ($Q$ and $a$) and the impact parameter $b=J/E=\pm K/E$ (in the equatorial plane).
It is easy to see that:
\be
{1\over\hat{r}^2}\mathcal{R}(\hat{r}) = \left({P_{\hat{r}}\hat{r}\over E}\right)^2={1\over\hat{r}^2} {\prod_{i=1}^4(\hat{r}-\hat{r}_i)}
\ee
is invariant under generalised CT inversion 
\be
\hat{r}'={\hat{r}_c^2\over \hat{r}}
\ee
since  ${\hat{r}_c^2/\hat{r}_{1,2}}=\hat{r}_{2,1}$ and ${\hat{r}_c^2/\hat{r}_{3,4}}=\hat{r}_{4,3}$.
The fixed locus of the inversion is 
\be
\hat{r} = \hat{r}_c = \sqrt{2a^2-a b+Q^2} 
\ee
that depends on the impact parameter $b$ and coincides with the photon-sphere after setting $b=b_c$. Indeed $\hat{r}_c$ is the radius of one of the two null critical geodesics satisfying 
\be\label{RR'}
\mathcal{R}(\hat{r}_c) = 0 = \mathcal{R}'(\hat{r}_c)
\ee
In the equatorial plane, for co-rotating geodesics ($ab>0$) one finds 
$$
\hat{r}^-_c=\max\{0,M-2|a|\}\quad,\quad b_c^-=\max\{2M,4(M-|a|)\}
$$
while for counter-rotating geodesics ($ab<0$) 
$$
\hat{r}^+_c=M+2|a|\quad,\quad b_c^+=4(M+2|a|)
$$
We start to see the emergence of the photon-halo ... but before dwelling on that, let us consider the characteristic function $S$ that encodes all the observables associated to (null) geodesics, such as scattering or in falling angles and Shapiro time delay.

For separable axisymmetric systems like KN BHs it reads 
\be\label{HamJak}
S=-Et+J\phi+S_r + S_\theta
\ee
with  
\be
S_r = \int_{r_i}^{r_f} P_r dr 
\ee
and 
\be
S_\theta = \int_{\theta_i}^{\theta_f} P_\q d\q
\ee
For geodesics in the equatorial plane $S_\theta=0$ and the only non-trivial part is the radial action $S_r$. Since $P_r= P_{\hat{r}} \approx \pm E$ at large distances, $S_r$
diverges when the initial or final points are taken to infinity. A simple regulator is to replace $r_f\rightarrow \infty$ with $r_f=\Lambda >> a,M,Q$. A similar problem emerges when the initial or final points are taken to the horizon $\hat{r}\rightarrow 0$.

The `regulated' radial action for scattering from infinity reads:
\be\label{regscattaction}
S_r^{scatt,reg}(E,J,\hat{r}_1,\Lambda)=\int_{\hat{r}_1}^{\Lambda}P_{\hat{r}}d\hat{r}=E\int_{\hat{r}_1}^{\Lambda}{\sqrt{\prod_{i=1}^4(\hat{r}-\hat{r}_i)}}{d\hat{r}\over\hat{r}^2}
\ee
Under CT inversion it transforms into the `regulated' radial action for falling into the horizon with the same energy $E$ and angular momentum $J$
\be\label{regfallaction}
S_r^{fall,reg}(E,J,\hat{r}_c^2/\Lambda,\hat{r}_2)=E\int_{\hat{r}_c^2/\Lambda}^{\hat{r}_2}{\sqrt{\prod_{i=1}^4(u-\hat{r}_i)}\over u^2}du=\int_{\hat{r}_c^2/\Lambda}^{\hat{r}_2}P_u du.
\ee
in so far as motion approaches the horizon within a distance $\varepsilon = \hat{r}_c^2/\Lambda$ that tends to zero when the cutoff $\Lambda$ is taken to infinity.
 
The `regulated' deflection angle
\be
\Delta\phi^{reg}(r_i,r_f)={\partial S^{reg}_r(E,J,r_i,r_j)\over\partial J},
\ee
can be computed by means of \eqref{regscattaction} and \eqref{regfallaction}.  Since the end-points of the `regulated' integration range depend on $J$ after CT inversion, the derivative of $S_r(E,J,r_i(J),r_f(J))$ w.r.t. $J$ involves a boundary term\footnote{We thank Alfredo Grillo for discussions on this point.}:
$${\partial S_r(E,J,r_i(J),r_f(J))\over\partial J}=\int_{r_i(J)}^{r_f(J)}{\partial P_r(r,E,J)\over\partial J}d r+{\partial r_f(J)\over\partial J}P_r(r_f(J),E,J)-{\partial r_i(J)\over \partial J}P_r(r_i(j),E,J)$$
As a consequence, one finds 
\be
\Delta \phi^{reg}_{scatt}=\int_{\hat{r}_1}^{\Lambda}{\partial P_{\hat{r}}\over\partial J}d\hat{r}
\ee
\be
\Delta \phi^{reg}_{fall}=\int_{\hat{r}_c^2(J)/\Lambda}^{\hat{r}_2}{\partial P_{\hat{r}}\over\partial J}d\hat{r}-\Lambda\left({\partial \hat{r}_c^2(J)\over \partial J}\right){E\over \hat{r}_c^4}\sqrt{\prod_{i=1}^4\left({\hat{r}_c^2(J)\over \Lambda}-\hat{r}_i\right)}
\ee
We conclude that the CT inversion symmetry holds `only' at the level of the `regulated' radial actions, whereby $S_r^{scatt,reg}$ transforms into $S_r^{fall,reg}$ and {\it vice versa}, while equality between scattering and falling angles is lost to some extent, due the dependence on $J$ of the product of the cutoffs $\varepsilon \Lambda = \hat{r}_c^2(J)$. Notice the analogies and the differences with the B2B formula \cite{Kalin:2019rwq, Kalin:2019inp} whereby an analytic continuation is needed.

\subsection{Non-planar Motion, non-equatorial geodesics}

When $P_\theta = \rho^2 \dot\theta \neq 0 $ the angular dynamics becomes highly non-trivial. Yet, thanks to separability the system is integrable and one can write 
Hamilton principal function as in \eqref{HamJak}.
%\be
%S=-Et+\int P_{\hat{r}}d\hat{r}+\int P_\q d\q+J\phi 
%\ee
The angular part of the action is quite non-trivial and reads
\be
S_\theta = \int_{\theta_0}^{\theta_f} P_\theta d\theta = E \int_{\theta_0}^{\theta_f}
\sqrt{b^2\sin^2\q-\left(\z+a\cos^2\q\right)^2} 
{d\theta \over \sin\theta} = \ee
$$ = E \int_{\chi_0}^{\chi_f} {d\chi \over 1-\chi^2}
\sqrt{b^2(1-\chi^2)-\left(\z+a\chi^2\right)^2}
$$
with $\chi=\cos\theta$, 
but will play only a marginal role in the following. 

Let us focus on the radial part. In general the four roots of ${\cal R}$ are 
\be
\hat{r}^{[\pm]}_{\pm}=-\sqrt{a^2+Q^2} +{1\over 2}\left[\pm b [\pm]\sqrt{b^2 +4 a\zeta \mp 4b\sqrt{a^2+Q^2}}\right]
\ee
The generalised CT inversion $\hat{r} \rightarrow \hat{r}_c^2/\hat{r}$ with 
\be
\hat{r}_c^2 = M^2-a\zeta = 2a^2+ Q^2 - a {J\over E}  
\ee
formally leaves the radial action invariant, up to regularisation. Due to frame dragging
$\hat{r}_c$ depends on the impact parameters $\zeta$ and $b$. The conditions for criticality $\mathcal{R}(\hat{r}_c) = 0 = \mathcal{R}'(\hat{r}_c)$ are easier to solve for $\zeta_c$ and $b_c$ in terms of $\hat{r}_c$ and yield
\be 
a\zeta_c = M^2 - \hat{r}^2_c \quad , \quad b_c^2 = 4 (\hat{r}_c + M)^2
\ee
with $M=M_e= \sqrt{a^2+Q^2}$ for extremal KN BHs. 
 
In Appendix \ref{appA} we (re)derive the allowed range for $r_c$ for (non)extremal KN BHs that characterizes the photon halo. In the extremal case of interest here, it turns out be $$
{\rm Max}\{0, M_e-2a\} \le \hat{r}_c \le M_e+2a$$
Correspondingly 
$$
{\rm Max}\{2M_e, 4(M_e-a)\} \le b_c \le 4(M_e+a)$$
with $|\zeta_c| \le b_c$, that is saturated on the equatorial plane as we have seen before. Notice that for $a>Q/\sqrt{3}$ (whereby $M_e<2a$) the inner critical geodesics lies on the horizon. This plays a role in the phenomenon of super-radiance that reflects in the presence of special (near) zero-damping modes to which we will come back later on. 

In order to visualize the photon-halo,  recall that for non planar motion $b=K/E$ and $\zeta= J/E - a$ are related by 
\be 
\sin^2\q_0 P_{\q,0}^2  ={E^2}[b^2\sin^2\q_0-\left(\z+a\cos^2\q_0\right)^2]
\ee
for an observer at infinity on a plane at fixed $\theta$, the collection of all points leading to a critical geodesics delimits the edge or rim of the BH
shadow. This is encoded in the parametric curve 
\be 
Y^2 + \left(X-X_0\right)^2 =b_c^2
\ee
with $Y= P_\theta(\hat{r}_c) /E$, $X= J(\hat{r}_c)/E\sin\theta$ and $X_0= a\sin\theta$ for $\hat{r}_c$ varying in the (sub)interval allowed by the chosen value of $\chi=\cos\theta$. See for instance \cite{Bianchi:2020des, Bianchi:2020yzr, Bacchini:2021fig} for further details and the connection with chaos and Lyapunov exponent.

\section{Extremal Rotating BHs in STU supergravity}\label{eSTURBH}

In this Section, we extend the previous analysis of CT transformations to eSTURBHs (extremal rotating BHs  in STU supergravity) \cite{Cvetic:1996xz, Cvetic:2020kwf, Cvetic:2021lss, Cvetic:2021vxa}. For simplicity we only consider charged BHs with 4 electric charges that can be obtained adding angular momentum to bound states of 4 stacks of D3-branes wrapped around (intersecting) 3-cycles in a torus $T^6$ or a CY 3-fold. 
\subsection{Geodesic motion and CT inversion}
Parameterizing the 4 charges as $Q_i = 2ms_ic_i$ with $m$ a mass scale, $s_i=\sinh\delta_i$ and  $c_i=\cosh\delta_i$ and neglecting scalar and abelian vector fields, that play no role in our analysis, the metric reads \cite{Cvetic:2020kwf, Cvetic:2021lss}  
\be\label{cpsbh}
ds^2 =-{\rho^2-2m r\over W}\left(dt+\mathcal{B}_{(1)}d\phi\right)^2+W\left({dr^2\over\Delta}+d\q^2+{\Delta\sin^2\q d\phi^2\over\rho^2-2mr}\right),
\ee
with
\be\mathcal{B}_{(1)}=2 m a \sin^2\q{ r\prod_c-(r-2m)\prod_s\over\rho^2-2m r},\quad\rho^2=r^2+a^2\cos^2\q,\quad\Delta=r^2-2m r+a^2,
\ee
$$W^2=\prod_{i}R_i+a^4\cos^4\q\nn
+2 a^2\cos^2\q\left[r^2+m r\sum_is_i^2+4m^2\left(\prod_c-\prod_s\right)\prod_s-2m^2\sum_{i<j<k}s_i^2s_j^2s_k^2\right],$$
$$R_i=r+2ms_i^2,\qquad\prod_c=\prod_ic_i,\qquad\prod_s=\prod_is_i.$$
It is straightforward but tedious to check that for $Q_1{=}Q_2{=}Q_3{=}Q_4{=}Q$ \eqref{cpsbh} reproduces the KN metric. The details are relegated in an Appendix. Let us stress that the mass parameter $m$ is not precisely the BH mass. The ADM mass is actually: $M={1\over4}\sum_i\sqrt{m^2+Q_i^2}$.

The condition for extremality is
\be
m=a 
\ee 
so that $\rho^2-2m r= (r-a)^2 - a^2\sin^2\theta$ and $\Delta = (r-a)^2$ and the event horizon is then at $r=a$. Setting 
\be
\hat{r}=r-a
\ee
the horizon is located at $\hat{r}=0$.  

For neutral massless probes, the null mass-shell condition in Hamiltonian form reads
\be 
\hat{r}^2P_{\hat{r}}^2+P_{\q}^2+{\hat{r}^2{-}a^2\sin^2\q\over \hat{r}^2\sin^2\theta}\left\{(J+E \mathcal{B}_{(1)})^2-{\hat{r}^2\sin^2\theta W E^2 \over(\hat{r}^2{-}a^2\sin^2\q)^2}\right\}=0
\ee
%where
%\be
%V(r;J,E) = ... \quad \Lambda(\theta;J,E) = ...
%\ee
%Introduction the analogue of Carter's separation constant one has
%\be
%P_\theta^2 + \Lambda(\theta;J,E) = K^2
%\ee 
%and
%\be
%\Delta_r P_r^2 - F(r;J,E) = - K^2
%\ee
%
%In the extremal case ($M=a$!!??)  $\Delta_r = (r-a)^2$ so that setting $\hat{r}=r-a$ the horizon is located at $\hat{r}_H=0$. Writing 
The separation is straightforward and yields
\be
{P_\q^2\over E^2}+{b^2\over \sin^2\q}+a^2\sin^2\q=\lambda^2
\ee
$${\hat{r}^2P_{\hat{r}}^2\over E^2}={\mathcal{R}(\hat{r})\over \hat{r}^2}$$
with $\lambda^2$ a positive constant and ${\cal R}(\hat{r})$ a quartic polynomial {\it viz.}
\begin{align}
{\cal R}(\hat{r}) &= \prod_i (\hat{r}-\hat{r}_i)= \hat{r}^4+2a\left(2+\sum_is_i^2\right)\hat{r}^3+\left[2a^2(4+3\sum_is_i^2+2\sum_{i<j}s_i^2s_j^2)-\lambda^2\right]\hat{r}^2+ \nn\\
&+\left[8a^3(1+\sum_is_i^2+\sum_{i<j}s_i^2s_j^2+\sum_{i<j<k}s_i^2s_j^2s_k^2)-4a^2b\left(\prod_c-\prod_s\right)\right]\hat{r}+\\\nn
&+a^2b^2-4a^3 b(\prod_c+\prod_s)+4a^4\left[1+\sum_is_i^2+\sum_{i<j}s_i^2s_j^2+\sum_{i<j<k}s_i^2s_j^2s_k^2+2\left(\prod_c+\prod_s\right)\prod_s\right] 
\end{align}
where $b=J/E$. Very much as for massless neutral scalar waves \cite{Cvetic:2020kwf, Cvetic:2021lss}, symmetry of radial motion under generalised CT inversions $\hat{r} \rightarrow \hat{r}_c^2/\hat{r}$ at fixed charges requires 
\be
\hat{r}_1\hat{r}_2=\hat{r}_3\hat{r}_4=\hat{r}_c^2(b,\lambda) 
\ee
where $\hat{r}_1,\hat{r}_2$ are the positive roots and $\hat{r}_3,\hat{r}_4$ are the negative ones and 
\be\label{6122}
\hat{r}_c^2(b,\lambda) = 2a^2\left(\prod_c+\prod_s\right)-ab 
\ee
that follows from the remarkable identity
\be
\left[1+\sum_is_i^2+\sum_{i<j}s_i^2s_j^2+\sum_{i<j<k}s_i^2s_j^2s_k^2+2\left(\prod_c+\prod_s\right)\prod_s\right] = \left(\prod_c+\prod_s\right)^2 
\ee
where the sign in \eqref{6122} is chosen by comparing with $r_c$ in Kerr. This is possible iff the charges $Q_i$ or equivalently the ``boost" parameters $\delta_i$ satisfy the condition 
\be
\tilde{\delta}_i \equiv {1\over 2} \sum_j\delta_j -\delta_i = \delta_{\pi(i)}  
\ee
with $\pi$ a permutation of the four indices, that is allowed since the metric is permutation invariant. Taking a pair of indices $i$ and $\pi(i)$ to be $1,2$ without loss of generality, one has
\be\label{cond-delta}
2\delta_1 + 2\delta_2 = \sum_j \delta_j = {1\over 2} \sum_j \log\left(q_j+\sqrt{q_j^2+1}\right)
\ee
with $q_j = Q_j/m = Q_j/a$,  that is satisfied when
\be\label{chargeCTcond}
\left(Q_1+\sqrt{Q_1^2+a^2}\right)\left(Q_2+\sqrt{Q_2^2+a^2}\right)= \left(Q_3+\sqrt{Q_3^2+a^2}\right)\left(Q_4+\sqrt{Q_4^2+a^2}\right)
\ee
or permutations thereof. This condition generalises in an interesting way the condition 
$Q_1Q_2 = Q_3Q_4$ found in \cite{Bianchi:2021yqs} for extremal non-rotating STU BHs\footnote{In STU supergravity, the permutation of the charges is an $SL(2,Z)^3$ U-duality transformation that leaves the metric invariant by definition. CT invariance selects (positive) charge configurations that are invariant under U-duality combined with Freudenthal duality $Q_i'= \sqrt{\prod_j Q_j / Q_i}$ \cite{Borsten:2018djw, Borsten:2019xas}. We thank Guillaume Bossard for clarifying comments on this issue.} . The special case $Q_1=Q_2=Q_3=Q_4=Q$, that we have shown to coincide with KN or, for $Q=0$, with Kerr, and the special cases with pairwise equal charges, e.g. $Q_1=Q_3$ and $Q_2=Q_4$, are simply sub-cases of the general case, that survive the inclusion of angular momentum $a$.  

Although, integrality of charges and spin may not be an issue for large `astrophysical' BHs, it is amusing to observe that a particularly simple integer solution of \eqref{chargeCTcond} is $Q_1=18$, $Q_2=32$, $Q_3=10$, $Q_4=45$, $a=24$.
It is also easy to convince oneself that \eqref{chargeCTcond} admit an infinite number of integer solutions. 

For 4-charge extremal rotating BHs, satisfying the above condition and thus admitting a proper (generalised) CT symmetry, the analysis of the observables proceeds along the same lines as in the simpler KN contexts. The radial action is form-invariant. Null infinity is exchanged with the horizon. Since the metric is not invariant, even if one allows Weyl rescalings, it does not make much sense to ask how the ergo-region transforms. Yet it is quite remarkable that the photon-halo is left fixed.

In the equatorial plane ($\theta=\pi/2$), the scattering angle $\Delta\phi_{scatt}(E,J)$ is mapped into the (regulated) in spiraling angle $\Delta\phi_{scatt}(E,J)$ for massless neutral probes with the same energy $E$ and angular momentum $J$. While the former is finite for non-critical values, the latter diverges due to the $\hat{r}^2$ factor in the denominator. In fact the divergence may be imputed to a boundary contribution generated by the dependence on $J$ or $b$ of the transformed extrema $\hat{r}' = \hat{r}_c^2(J,E)/\hat{r}$. 

Since geodesics are generically non-planar as in the KN case, the relevant observable is the full action, including the non-trivial angular part $S_\theta$, but the only part acted on by CT transformations is the radial action $S_r$. Even though conformal spatial inversions $\vec{x}' = - r_c^2 \vec{x}' / |\vec{x}|^2$ seem to produce an antipodal transformation $\theta\rightarrow \pi-\theta$ ($\chi=\cos\theta \rightarrow -\chi$) the metric and the geodesics are invariant under and the angular action $ \int P_\theta d\theta$ is unaffected, up to exchange of the extrema of integration.

\subsection{Critical regime}

Once shown that massless geodesics in eSTURBH's metric \eqref{cpsbh} admit CT conformal inversion as symmetries, let us focus on the critical geodesics that form the halo, fixed under CT transformations. 

Let us step back and reconsider the non extremal case. Later on we will refocus on the extremal case.

In general for $m\neq a$, the zero mass shell condition in Hamiltonian form can be written as
\begin{align}
&\Delta P_r^2-{a^2 J^2\over \Delta}+{2ma[r\prod_c-(r-2m)\prod_s]\over\Delta}2 E J-{E^2\over2\Delta}\Biggl\{a^4+2\prod_{i=1}^4R_i+a^2\Biggl[3r^2+2mr\left(1+2\sum_{i=1}^4\sigma_i^2\right)+\\\nn
&-8m^2\left(\sum_{i<j<k}\sigma_i^2\sigma_j^2\sigma_k^2+2{\prod_s}^2-2\prod_c\prod_s\right)\Biggl]-a^2\Delta\Biggl\}+P_\q^2+{J^2\over\sin^2\q}-E^2a^2\cos^2\q=0
\end{align}
If we introduce the separation constant $\lambda^2$, we obtain:
\begin{align}
{P_\q^2\over E^2}&=\lambda^2-{b_J^2\over\sin^2\q}-a^2\sin^2\q\\\nn
{\Delta P_r^2\over E^2}&=-\lambda^2+{a^2 b_J^2\over \Delta}-{4mab_J[r\prod_c-(r-2m)\prod_s]\over\Delta}+{1\over2\Delta}\Biggl\{a^4+2\prod_{i=1}^4R_i+\\\nn&+a^2\Biggl[3r^2+2mr\left(1+2\sum_{i=1}^4\sigma_i^2\right)
-8m^2\left(\sum_{i<j<k}\sigma_i^2\sigma_j^2\sigma_k^2+2\left(\prod_s\right)^2-2\prod_c\prod_s\right)\Biggl]+{a^2\Delta\over 2}\Biggl\}
\end{align}
where $\lambda=K/E$ and $b_J=J/E$. For brevity we set
\begin{align}
&\mathcal{R}_4(r)={\Delta^2P_r^2\over E^2}=r^4+Ar^3+Br^2+Cr+D-\Delta\lambda^2\\\nn
&A=2m\sum_{i=4}^4\sigma_i^2\quad,\quad B=2a^2+4m^2\sum_{i<j}\sigma_i^2\sigma_j^2\\\nn
&C=C_1 b_J+C_2\quad,\quad C_1=4ma\left(\prod_s-\prod_c\right)\quad,\quad C_2=2a^2 m\sum_{i=1}^4\sigma_i^2+8m^3\sum_{i<j<k}\sigma_i^2\sigma_j^2\sigma_k^2\\\nn
&D=a^2 b_J^2-8m^2a\prod_s b_J+D_1\quad,\quad D_1=a^4+16m^4\prod_s{}^2-4a^2m^2\left(\sum_{i<j<k}\sigma_i^2\sigma_j^2\sigma_k^2+2\prod_s{}^2-2\prod_s\prod_c\right)
\end{align}
In the critical regime we have:
\begin{align}
&r^4+Ar^3+Br^2+Cr+D=\Delta \lambda^2\\\nn
&4r^3+3Ar^2+2Br+C=2(r-m)\lambda^2
\end{align}
The solutions to this system are
\begin{align}\label{criticalsolution}
&b_J={m\over a(r-m)}\Biggl[(a-2m+r)(a+2m-r)\prod_s+(r-a)(r+a)\prod_c-{\sqrt{\tilde{\Delta}}\over4am}\Biggl]\\\nn 
&\tilde{\Delta}=16a^2\Delta^2\Biggl[r^2+m\sum_{i=1}^4\sigma_i^2 r+m^2\left(\left(\prod_s-\prod_c\right)^2-1-\sum_{i=1}^4\sigma_i^2\right)\Biggl]
\end{align}
and
\be
\lambda^2={4r^3+3Ar^2+2Br+C_2\over2(r-m)}+{C_1\over2(r-m)}b_J
\ee
where for $b_J$ we choose the negative sign in order to match with KN. In order to match with the angular equation in \eqref{ptheta}, we have to redefine the angular momenta as follows:
\be
b^2=\lambda^2-2ab_J\quad,\quad\z=b_J-a
\ee
Thanks to the condition $b\ge\z$ descending from the non-negativity of \eqref{ptheta}, we can identify the photon region. 
%\begin{figure}[h!]
%\centering
%\includegraphics[width=0.9\textwidth]{cpsbz}
%\caption{The plots of b and $|\z|$ for $a=0.8$, $m=1$, $Q_i=i$, $i=1,...,4$}\label{fig10}
%\end{figure}

In  the extremal limit, i.e. when $m=a$, we have
\begin{align}\label{b&z}
&\z=-a+(3a-r)\prod_s+(a+r)\prod_c+{a-r\over a}\sqrt{r^2+a\sum_{i=1}^4\sigma_i^2 r+a^2\left(-1-\sum_{i=1}^4\sigma_i^2+\left(\prod_s-\prod_c\right)^2\right)}\\\nn
&b^2=2r^2+a\left(2+3\sum_{i=1}^4\sigma_i^2\right)r+a^2\left(2+\sum_{i=1}^4\sigma_i^2+2\sum_{i<j}\sigma_i^2\sigma_j^2-2\sum_{i<j<k}\sigma_i^2\sigma_j^2\sigma_k^2-4{\prod_s}^2+4\prod_s\prod_c\right)+\\\nn
&-2a(3a-r)\prod_s-2a(a+r)\prod_c-2\Biggl[a-r+a\left(\prod_s-\prod_c\right)\Biggl]\times\\\nn
&\times\sqrt{r^2+a r \sum_{i=1}^4\sigma_i^2-a^2\left[1+\sum_{i=1}^4\sigma_i^2-\left(\prod_s-\prod_c\right)^2\right]}
\end{align}
%\begin{figure}[h!]
%\centering
%\includegraphics[width=0.9\textwidth]{cpsbze}
%\caption{The plots of b and $|\z|$ for, $a=m=1$, $Q_i=i$, $i=1,...,4$}\label{fig11}
%\end{figure}
If we take the charges s.t. $\sigma_1=\sigma_2=\sigma$ and $\sigma_3=\sigma_4=\tau$, \eqref{b&z} drastically simplifies
\begin{align}
&\z=-{1\over a}[(r-a)^2-a^2(1+2\sigma^2)(1+2\tau^2)]\\\nn
&b^2=4[r+a(\sigma^2+\tau^2)]^2
\end{align}
The condition for $r_c^{-}$ to be outside the horizon $r_c^{-}\ge r_H=a$ boils down to\be
b(r_H)\le\z(r_H)\Longrightarrow \sigma^2\tau^2\ge{1\over 4}
\ee
%\begin{figure}[h!]
%\centering
%\includegraphics[width=.4\textwidth]{sigmatauminore14}\hfil
%\includegraphics[width=.4\textwidth]{sigmatauuguale14}\hfil
%\includegraphics[width=.4\textwidth]{sigmataumaggiore14}
%\caption{In clockwise sense from the upper figure on the left, there are the plots of b and $|\z|$ for $\sigma^2\tau^2=1/10<1/4$, then $\sigma^2\tau^2=1/4$ and finally $\sigma^2\tau^2=1/2>1/4$}\label{fig11}
%\end{figure}
In the case of four different charges, the condition $r_c^-\ge r_H$ becomes
\be\label{*}
3+2\sum_{i=1}^4\sigma_i^2-4\sum_{i<j<k}\sigma_i^2\sigma_j^2\sigma_k^2-8\prod_s\left(\prod_c+\prod_s\right)\le 0
\ee
If the condition for invariance under CT inversions, that can be written as:
\be\label{CT}
2+\sum_{i=1}^4\sigma_i^2=2\left(\prod_c-\prod_s\right)
\ee
is obeyed, the expression for $\z$ (and consequently for $\lambda^2$) can be simplified to
\be
\z={-\Delta-a^2+2m^2(\prod_c+\prod_s)\over a}+{2m(m^2-a^2)(\prod_c-\prod_s)\over a(r-m)}
\ee
... another bonus of CT invariance.
\section{Waves in 4 charge STU BH background and Super-radiant modes}\label{STUBHwave}

In this section we switch gear and consider the so-called (near) super-radiant modes of (near-estremal) STURBHs. The photon-sphere, that is the fixed locus of CT inversions if allowed, plays a crucial role in the linear response of BHs and other compact objects to small perturbations. In particular the ring-down phase of BH mergers is known to be dominated by Quasi Normal Modes (QNMs). These are fluctuations that satisfy outgoing b.c. at infinity and ingoing b.c. at the horizon. In the WKB approximation the frequency is given by $$\omega^{WKB}_{QNM} = \omega_c - i \lambda (2n + 1)$$ where $ \omega_c$ is the frequency of critical circular null orbits, while $\lambda$ is the Lyapunov exponent governing the chaotic behaviour of nearby critical geodesics
\cite{Bianchi:2020des, Bianchi:2020yzr, Bacchini:2021fig}. A detailed study of the full spectrum of QNMs of STURBHs is beyond the scope of the present investigation and we hope to report on this soon. 
Exact results may be obtained resorting to the surprising connection between 
QNMs and quantum SW curves of ${\cal N}=2$ SYM with gauge group $SU(2)$ and $N_f$ hypermultiplets in the fundamental, that we will exploit later on to some extent. 

Near extremal BHs however possess a special class of QNMs: near super-radiant modes, also known as zero-damping modes (ZDMs), since ${\rm Im}\omega_{ZDM}$ is very small and vanishes in the extremal limit. These modes are produced by near extremal BH mergers and thanks to their slow fall-off in time provide a very peculiar feature of the ring-down phase in these cases. The super-radiant threshold frequency turns out to be 
\be
\omega_{SR}={m_\f}\Omega_H 
\ee
where $\Omega_H$ is the angular velocity at the horizon. Near super-radiant modes are defined by taking
\be
\omega=\omega_{SR}+\n\d
\ee
with finite $\nu=\nu_1+i\nu_2$ to be determined and $\d=r_+-r_-<<r_{\pm}$, the small separation between the inner and the outer horizon.

The super-radiant threshold frequency for NESTURBHs (near-extremal STU rotating BHs) is given by the above expression with
\be 
\Omega_H=-{g_{t\phi}(r_+)\over g_{\phi\phi}(r_+)}=-{1\over \mathcal{B}_{(1)}(r_+)}={a\over2m(r_+\prod_c-(r_+-2m)\prod_s)}
\ee 
Since the horizons are located at
\be
r_\pm=m\pm\sqrt{m^2-a^2}
\ee
setting
\be
m^2=a^2+{\delta^2\over 4}
\ee
at zero order in $\delta$, one has
\be
r_+\simeq m\simeq a,\quad,\quad \Omega_H\simeq{1\over 2a\left(\prod_c+\prod_s\right)}
\ee
Starting from \eqref{cpsbh}, it is straightforward to separate variables in the massless scalar wave equation $\Box \Phi=0$ in this background. Setting
\be
\Phi=e^{-i\omega t+im\phi}{\psi(r)S(\chi)\over\sqrt{\Delta (1-\chi^2)}}\quad,\quad \chi=\cos\q
\ee
and introducing the separation constant $\lambda^2$ bring the two wave equations into Sch\"odinger-like canonical form.
The angular equation determines the spheroidal harmonics 
\be\label{eqdiff}
S_{\lambda,m_\f}''(\chi)+{(1-\chi^2)(a^2\omega^2\chi^2-a^2\omega^2+\lambda^2)+1-m_\f^2\over(1-\chi^2)^2}S_{\lambda,m_\f}(\chi)=0
\ee
For small $a^2\omega^2$,  $\lambda^2=\ell(\ell+1) + {\cal O} (a^2\omega^2)$ and $S_{\ell,m}(\chi) = P_{\ell,m}(\chi) + {\cal O} (a^2\omega^2)$ are known as spheroidal harmonics. The problem can be solved using standard perturbation theory in Quantum Mechanics, even though the wave equation is classical, or  identifying the (confluent) HE \eqref{eqdiff}
(with two regular and one irregular singularities) with the quantum SW curves for ${\cal N}=2$ SYM with gauge group $SU(2)$ and $N_f=3$ hypermultiplets in the fundamental (six doublets) \cite{Aminov:2020yma, Bianchi:2021xpr, Bonelli:2021uvf, Bianchi:2021mft, Bonelli:2022ten}.

If we define $A=\lambda^2-a^2\omega^2$ as in \cite{Bianchi:2021mft}, the radial equation can be written as:
\begin{align}
\psi''&+\Biggl\{\omega^2\Delta\Biggl[r^2+2m\left(1+\sum_i\sigma_i^2\right)r-4m^2\left(\sum_{i<j<k}\sigma^2\sigma^2\sigma^2-2\prod_s\left(\prod_c-\prod_s\right)\right)\Biggl]+{\Delta'^2\over 4}-\Delta+\\\nn
&+4m^2\left(\omega_{SR}\Biggl[r_+\prod_c-(r_+-2m)\prod_s\Biggl]-\omega\Biggl[r\prod_c-(r-2m)\prod_s\Biggl]\right)^2-\Delta A\Biggl\}{\psi\over\Delta^2}=0
\end{align}
%Let's introduce new variables in order to rewrite the differential equations \eqref{eqdiff} in canonical form:
%\be
%R(\hat{r})={\psi(\hat{r})\over\hat{r}}\quad,\quad S(y)={Z(y)\over\sqrt{1-y^2}}.
%\ee
%So \eqref{eqdiff} become:
%\begin{align}
%&\psi''+Q_{\hat{r}}\psi=0\quad,\quad Z''+Q_\q Z=0\\\nn
%&Q_{\hat{r}}={H-\lambda^2\over\hat{r}^2}\\\nn
%&Q_\q={(1-y^2)[a^2\omega^2y^2-a^2\omega^2+\lambda^2]-m_\phi^2+1\over(1-y^2)^2}
%\end{align}
%In order to get contact with the analogue results for KN BH (in BCGM), we define $A=\lambda^2-a^2\omega^2$.
Far from the horizon $r\gg r_+\gg\delta$ the radial equation reduces to:
\be
\psi''(r)+\Biggl[\omega_{SR}^2+{2r_+\omega_{SR}^2(2+\sum_i\sigma_i^2)\over(r-r_+)}+{r_+^2\omega_{SR}^2\left(7+6\sum_i\sigma_i^2+4\sum_{i<j}\sigma_i^2\sigma_j^2\right)-A\over (r-r_+)^2}\Biggl]\psi(r)=0
\ee
Requiring $\psi\sim e^{i\omega_{SR}\hat{r}}$ at infinity (outgoing) and as $\psi\sim\hat{r}^{1/2+\a}$ at the horizon (regularity) one has 
\be\label{infinity1}
\psi(r)= c_\infty e^{i\omega_{SR}(r-r_+)} (r-r_+)^{1/2+\a}U(\tilde{A},\tilde{B};z)
\ee
where $c_\infty$ is a constant and $U$ is Tricomi confluent Hypergeometric function, while
\begin{align}\label{infinity}
&z=-2 i \omega_{SR}(r-r_+)\\\nn
&\a^2=A+{1\over4}-a^2\omega_{SR}^2\left(7+6\sum_i\sigma_i^2+4\sum_{i<j}\sigma_i^2\sigma_j^2\right)\\\nn
&\tilde{A}=\a+{1\over2}- i \omega_{SR} a\left(2+\sum_i\sigma_i^2\right)\\\nn
&\tilde{B}=1+2\a
\end{align}
On the other hand, the radial equation in the near horizon limit can be approximated by defining $\tau = (r-r_+)/\delta$ with $\delta<<r_+$. In the variable $\t$ one finds
\be
\psi''(\tau)+Q(\t)\psi(\t)=0\ee
with
\be
Q(\t)={4a^2\Biggl[\omega_{SR}\t\left(\prod_c-\prod_s\right)+\n a\left(\prod_c+\prod_s\right)\Biggl]^2+{1\over 4}-\t(1+\t)\Biggl[\a^2+4\omega_{SR}^2a^2\left(\prod_c-\prod_s\right)^2-{1\over 4}\Biggl]\over\t^2(1+\t)^2}
\ee
The solution can be written in terms of hypergeometric functions. Imposing in-going boundary conditions at the horizon, one finds
\be\label{horizon1}
\psi(\t)=c_H\t^{{1\over 2}-{i a \n\over\Omega_H}}(1+\t)^{{1\over 2}-{i a \n\over\Omega_H}+2ia\omega_{SR}\left(\prod_c-\prod_s\right)}{}_2F_1(\bar{A},\bar{B},\bar{C};,-\t)
\ee
where $c_H$ is a constant and 
\begin{align}\label{horizon}
&\bar{A}={1\over 2}-\a-{2 i a \n\over\Omega_H}+2ia\omega_{SR}\left(\prod_c-\prod_s\right)\\\nn
&\bar{B}={1\over 2}+\a-{2 i a \n\over\Omega_H}+2ia\omega_{SR}\left(\prod_c-\prod_s\right)\\\nn
&\bar{C}=1-{2 i a \n\over\Omega_H}
\end{align}
By expanding \eqref{infinity1} near the horizon, one finds
\be\label{match1}
\psi(r)\sim(r-r_+)^{{1\over2}-\a}\Biggl[(-2i\omega)^{-2\a}{\Gamma(2\a)\over\Gamma(\tilde{A})}+{\Gamma(-2\a)\over\Gamma(\tilde{A}-\tilde{B}+1)}(r-r_+)^{2\a}\Biggl]
\ee
while far away from the horizon, \eqref{horizon1} reduces to
\be\label{match2}
\psi(r)\sim(r-r_+)^{{1\over 2}-\a}\Biggl[{\Gamma(2\a)\over\Gamma(\bar{B})\Gamma(\bar{C}-\bar{A})}(r-r_+)^{2\a}\delta^{-{1\over2}-\a}+{\Gamma(-2\a)\over\Gamma(\bar{A})\Gamma(\bar{C}-\bar{B})}\delta^{\a-{1\over2}}\Biggl]\Gamma(\bar{C})
\ee
Matching \eqref{match1} and \eqref{match2} one finds
\be
{\Gamma^2(2\a)\Gamma(\tilde{A}-\tilde{B}+1)\Gamma(\bar{A})\Gamma(\bar{C}-\bar{B})\over\Gamma(\bar{B})\Gamma(\bar{C}-\bar{A})\Gamma^2(-2\a)\Gamma(\tilde{A})}(-2i\omega_{SR}\delta)^{-2\a}=1.
\ee
Since $\delta\sim 0$ and $\Re\a>0$, the factor $\delta^{-\a}$ in the left hand side diverges, so it has to be compensated by a pole of $\Gamma(\bar{B})$ in the denominator\footnote{Since we want to determine the near super-radiance parameter $\nu$, we are forced to quantise $\bar{B}$. The arguments of the other   
$\Gamma$ functions in the denominator are independent of $\nu$. } i.e.
\be
\bar{B}=-n+(-2i\omega_{SR}\delta)^{2\a}\eta
\ee
with
\be
\eta={(-1)^n\over n!}{\Gamma^2(-2\a)\Gamma(\bar{C}-\bar{A})\Gamma(\tilde{A})\over\Gamma^2(2\a)\Gamma(\tilde{A}-\tilde{B}+1)\Gamma(\bar{A})\Gamma(\bar{C}-\bar{B})}
\ee
leading to
\be
\omega=\omega_{SR}+\delta\Biggl[{\Omega_H\omega_{SR}\left(\prod_c-\prod_s\right)-i{\Omega_H\over2a}(n+{1\over2}+\a)}\Biggl]+...
\ee
The imaginary part of the quasi-normal frequencies is expressed in terms of the Lyapunov exponent $\lambda_L = {\Omega_H\over2a}\approx T_{BH}$ \cite{Maldacena:2015waa, Bianchi:2020des, Bianchi:2020yzr} and the overtone number $n$ gets shifted by $\alpha + {1\over 2}$.

One might notice that imposing the condition for CT invariance \eqref{CT} allows to rewrite $\tilde{A}$ and $\tilde{B}$ in \eqref{infinity} in terms of $\bar{A}$, $\bar{B}$ and $\bar{C}$ in \eqref{horizon} so that
\be
\tilde{A}=\bar{C}-\bar{A}\quad,\quad\bar{C}-\bar{B}=\tilde{A}-\tilde{B}+1
\ee
This simplification is another bonus of the CT symmetry for NESTURBHs with special charges, satisfying  (\ref{chargeCTcond}).

\section{Rotating BHs and D3-branes in higher dimensions}\label{RotD3}

Very much as for non-rotating BHs and branes, one may ask whether rotating BHs and branes in higher dimensions enjoy invariance under generalised CT transformations. As we will see the answer is positive and we manage to identify at least one class of 5-d extremal BHs that enjoy this property. We will then address the issue in $D>5$ and we don't find any simple solution with this property. We do not exclude that more elaborate constructions with various scalars and gauge fields can enjoy CT invariance.

\subsection{$d=5$ CCLP solution}

%In string theory, 'large' (BPS) BHs with a finite area of the horizon require bound states of strings and branes with at least three charges. The three abelian charges couple to as many (gravi)photons, that in turn are dual to anti-symmetric tensors in five-dimensions, and to scalar fields parametrizing the shape and sizes of the internal manifold. One can ?identify? the three charges so as to reduce the number of ?active? (gravi)photons to one. In this setting scalar fields become constant and the relevant degrees of freedom are those of 

Five-dimensional charged rotating BH solutions to Einstein-Maxwell theory were found by Cheung, Cvetic, Lu and Pope (CCLP) in \cite{Chong:2004na, Chong:2005hr}. They depend on four parameters: mass $M$, charge $Q$ and two angular momenta $\ell_1$ and $\ell_2$. The metric reads 
\be
ds_5^2=-dt^2+\Delta_t(dt-\omega_1)^2-{2Q\over \rho^2}(dt-\omega_1)\omega_2+\rho^2\left(d\q^2+{dr^2\over\Delta_r}\right)+(r^2+l_1^2)\sin^2\q d\phi^2+(r^2+l_2^2)\cos^2\q\psi^2
\ee
while the abelian gauge field is
$$A={\sqrt{3}Q\over\rho^2}(dt-\omega_1)$$
where
\begin{align}
&\rho^2=r^2+l_1^2\cos^2\q+l_2^2\sin^2\q\\\nn
&\Delta_r={(r^2+l_1^2)(r^2+l_2^2)-2Mr^2+2l_1l_2Q+Q^2\over r^2}\quad,\quad\Delta_t={2M\over\rho^2}-{Q^2\over\rho^4}\\\nn
&\omega_1=l_1\sin^2\q d\phi+l_2\cos^2\q d \psi\quad,\quad\omega_2=l_2\sin^2\q d\phi+l_1\cos^2\q d\psi.
\end{align}
The curvature singularity of the metric is located at $\rho=0$, i.e.
$r^2_{sing}=-l_1^2\cos^2\q-l_2^2\sin^2\q$.
The horizons are located at 
\be\label{root}
r^2_{\pm}={1\over2}(2M-l_1^2-l_2^2)\pm{1\over2}\sqrt{(2M-l_1^2-l_2^2)^2-4(Q+l_1l_2)^2}
\ee
where $\Delta_r$ vanishes. There is also an ergo-surface that is delimited by the larger solution of $g_{tt}(r_{ergo})=0$ where $\Delta_t=1$ and $\rho^2_{ergo}=M\pm\sqrt{M^2-Q^2}$:
\be
r^2_{ergo}=M\pm\sqrt{M^2-Q^2}-l_1^2\cos^2\q-l_2^2\sin^2\q
\ee
Extremality ($T_{BH}=0$) requires $r_+=r_-$ that is
\be\label{CCLPextr}
|2M-l_1^2-l_2^2|=2|Q+l_1l_2|.
\ee
while supersymmetry (BPS condition) requires $Q=M$ \cite{Breckenridge:1996is}.

Geodesics of massless neutral gprobes can be written in Hamiltonian form $\mathcal{H}=0$ exploiting the conservation of three momenta: $P_t=-E$, $P_\phi=J_\phi$ and $P_\psi=J_\psi$. As in the KN and STURBH cases the Hamiltonian can be separated as
\be
\mathcal{H}={\mathcal{H}_r+\mathcal{H}_\q\over 2\rho^2}=0
\ee
with
\begin{align}
\mathcal{H}_r&=\Delta_r P_r^2+ {E^2\over r^2\Delta_r}[(r^2+\ell_1^2+\ell_2^2)(Q^2-r^2\Delta_r)-2M(r^2+\ell_1^2)(r^2+\ell_2^2)]+\\\nn
&+{P_{\psi}^2\over r^2\Delta_r}[(\ell_1^2-\ell_2^2)(r^2+\ell_1^2)-2\ell_1(M\ell_1+Q\ell_2)]+{P_{\phi}^2\over r^2\Delta_r}[(\ell_2^2-\ell_1^2)(r^2+\ell_2^2)-2\ell_2(M\ell_2+Q\ell_1)]+\\\nn
&+{2P_\psi E\over r^2\Delta_r}[(r^2+\ell_1^2)(2M\ell_2+Q\ell_1)-\ell_2(Q^2+r^2\Delta_r)]+{2P_\phi E\over r^2\Delta_r}[(r^2+\ell_2^2)(2M\ell_1+Q\ell_2)-\ell_1(Q^2+r^2\Delta_r)]
+\\\nn
&-{2P_\psi P_\phi\over r^2\Delta_r}[2M\ell_1\ell_2+Q(\ell_1^2{+}\ell_2^2)]
=-{K}^2\\\nn
\mathcal{H}_\q&=P_\q^2+\left(E\ell_1\sin\q+{P_\phi\over\sin\q}\right)^2+\left(E\ell_2\cos\q+{P_\psi\over\cos\q}\right)^2={K}^2
\end{align}
Once again, $K^2$ represents the square of the angular momentum. In a certain sense, one can then write the radial equation in the form
\be
\mathcal{R}(r)=\left({r\over 2E}{\partial\mathcal{H}_r\over\partial P_r}\right)^2={r^2\Delta_r^2P_r^2\over E^2}=Ar^6+Br^4+Cr^2+D
\ee
with
\begin{align}
A&=1\quad,\quad B=2\ell_2b_\psi+2\ell_1 b_\phi+2(\ell_1^2+\ell_2^2)-b^2\quad,\\\nn
C&=2Q\ell_1\ell_2+\ell_1^4+3\ell_1^2\ell_2^2+\ell_2^4+b^2(2M-\ell_1^2-\ell_2^2)+(\ell_1^2-\ell_2^2)(b_\phi^2-b_\psi^2)+\\\nn
&+2b_\phi(\ell_1(\ell_1^2+\ell_2^2-4M)-Q\ell_2)+2b_\psi(\ell_2(\ell_1^2+\ell_2^2-4M)-Q\ell_1)\\\nn
D&=(Q+\ell_1\ell_2)^2(-b^2+4b_\phi\ell_1+4b_\psi \ell_2)+2(Q+\ell_1\ell_2)(b_\phi^2\ell_1\ell_2+b_\psi^2\ell_1\ell_2+\ell_1^3\ell_2+\ell_1\ell_2^3+b_\phi b_\psi(\ell_1^2+\\\nn&+\ell_2^2)-b_\psi(\ell_1^3+2\ell_1\ell_2^2)-b_\phi(2\ell_1^2\ell_2+\ell_2^3))+(2M-\ell_1^2-\ell_2^2)(b_\psi^2\ell_1^2+2b_\phi b_\psi \ell_1\ell_2-2b_\psi\ell_1^2\ell_2+\\\nn&+b_\phi^2\ell_2^2-2b_\phi\ell_1\ell_2^2+\ell_1^2\ell_2^2)\end{align}
and
\be
b=\frac{K}{E}\quad,\quad b_\phi=\frac{P_\phi}{E}\quad,\quad b_\psi=\frac{P_\psi}{E}
\ee
are the impact parameters. Exchanging $\ell_1\leftrightarrow \ell_2$ corresponds to exchanging $b_\psi \leftrightarrow b_\phi$ and $\theta\leftrightarrow {\pi\over 2}-\theta$.

%For simplicity, we will focus on geodesic motion with constant $\q=\q_0$. This is only possible on the hyperplanes $\q=0,\pi/2$. For $\q_0=0$, the geodesics along these hyper-planes are characterised by the following impact parameters:
%\begin{align}
%&\q_0=0\quad,\quad b_\phi=0\quad,\quad b=b_\psi+\ell_2\\\nn
%&\q_0={\pi\over 2}\quad,\quad b_\psi=0\quad,\quad b=b_\phi+\ell_1
%\end{align}
%We will specialize to the plane $\q_0=0$. The result for $\q_0=\pi/2$ can be obtained from the ones presented here by 

Looking for extremal (not necessarily BPS) solutions  with $r_+=r_-=r_H$, that enjoy symmetry under CT inversions we further specialise to the case:
\be
2M=\ell_1^2+\ell_2^2\quad,\quad Q=-\ell_1 \ell_2. 
\ee
but we allow for $Q\neq M$ and $\ell_1\neq \ell_2$.
With this choice, the horizon falls onto the `origin' 
\be
r_{H}^2=0 \quad {\rm so \: that} \quad \Delta_r=r^2
\ee
and the parameters of the radial equation simplify to
\begin{align}
&A=1\quad,\quad B=2\ell_1 b_\phi+2\ell_2b_\psi+2(\ell_1^2+\ell_2^2)-b^2\\\nn
&C=(\ell_1^2-\ell_2^2)\Biggl[\left(b_\phi-{\ell_1^3\over\ell_1^2-\ell_2^2}\right)^2-\left(b_\psi-{\ell_2^3\over\ell_2^2-\ell_1^2}\right)^2\Biggl]\quad,\quad D=0
\end{align} 

The radial equation can be written as
\be
{\cal R}(r) = {r^4P_r^2\over E^2}=r^4+Br^2+C = (r^2-r_1^2) (r^2-r_2^2)
\ee
The zeros of $P_r$ are located at
\begin{align}
r_{1,2}^2={1\over 2}\Biggl\{b^2{-}2(\ell_1^2{+}\ell_2^2){-}2\ell_1 b_\phi{-}2\ell_2b_\psi{\pm}\Biggl[b^4{-}4b^2(\ell_1^2{+}\ell_2^2{+}\ell_1 b_\phi{+}\ell_2b_\psi){+}\\\nn{+}4[b_\phi^2\ell_2^2{+}b_\psi^2\ell_1^2{+}2b_\phi \ell_1(2\ell_1^2{+}\ell_2^2){+}2b_\psi \ell_2(2\ell_2^2{+}\ell_1^2) {-}4 \ell_1^2\ell_2^2\Biggl]^{1/2}\Biggl\}
\end{align}
so that:
\be\label{rootCCLP}
r_1r_2=\sqrt{C} = r_c^2(b_\psi^c,b_\phi^c) \quad {\rm with} \quad b_c^2(b_\psi^c,b_\phi^c)= 2\ell_1 b^c_\phi+2\ell_2b^c_\psi+2(\ell_1^2+\ell_2^2) + 2r_c^2
\ee
Indeed the criticality conditions ${\cal R}=0= {\cal R}'$ allow to express $r_c$ and $b^2_c$ in terms of $b_\psi^c,b_\phi^c$., so much so that `generalised' CT inversions that exchange horizon and infinity, keep the photon-halo $r=r_c(b_\psi^c,b_\phi^c)$ fixed, as expected. For consistency one has to impose positivity of $C=r_c^4$ and  that means
\be
\left(b_\phi-{\ell_1^3\over\ell_1^2-\ell_2^2}\right)^2> \left(b_\psi+{\ell_2^3\over\ell_1^2-\ell_2^2}\right)^2 \quad {\rm for } \quad \ell_1> \ell_2
\ee
or {\it vice versa} for $\ell_2> \ell_1$. Moreover $b_c^2\ge 0$ requires 
$$
\ell_1 b^c_\phi+\ell_2b^c_\psi> -\ell_1^2-\ell_2^2-r_c^2
$$

For $\ell_1=\pm \ell_2=\ell$ only the combination $b_\pm=b_\phi\pm b_\psi$ matters. 
In this case, choosing the plus sign and introducing $\hat{r}_c=r_c/\ell$ and $\hat{b}_{+,c}=b_{+,c}/\ell$, one has 
\be
\hat{r}_c^4= 3-2\hat{b}_{+,c} \quad , \quad \hat{b}_c^2 = 2 \hat{b}_{+,c}+4 + 2 \sqrt{3-2\hat{b}_{+,c}}
\ee
The minimal value for both $\hat{r}_c^2$ and $\hat{b}_c^2$ is zero. The former, $\hat{r}_c^2=0$ (horizon),  is reached for $\hat{b}_{+,c}= 3/2$ whereby  $\hat{b}_c^2=7$. The latter, $\hat{b}_c^2=0$, for $\hat{b}_{+,c}= -3-\sqrt{8}$, whereby $\hat{r}_c^2= 1+\sqrt{8}>\sqrt{3}$.

 For $\ell_1\neq\pm \ell_2$ the situation is much more involved. Although it is easy to determine the minimal value of $\hat{r}_c^2=0$ (horizon) that is reached for 
$b_\phi={\ell_1^3\over\ell_1^2-\ell_2^2}$ and $b_\psi= - {\ell_2^3\over\ell_1^2-\ell_2^2}$ whereby $b^2 = 4(\ell_1^2+\ell_2^2)$, the maximum is harder to determine. 
In fact, for $\ell_1^2>\ell_2^2$, taking $b_\psi= - {\ell_2^3\over\ell_1^2-\ell_2^2}$ and letting $\tilde{b}_\phi = b_\phi-{\ell_1^3\over\ell_1^2-\ell_2^2}$ grow arbitrarily, both $r_c^2$ and $b_c^2$ grow arbitrarily.

Thanks to the previous properties of the roots, the radial action is form-invariant
\be
S_r = \int P_r dr 
\ee
under CT inversions $r\rightarrow r_c^2/r$ that maps the scattering regime outside the photon-sphere into the in-spiraling regime inside it.

\subsection{Rotating D3 branes}

Let us now pass to consider rotating D3-branes.  Thanks to the transversal $SO(6)$ symmetry, one can introduce three indipendent angular momentum parameters $\ell_1$,$\ell_2$ and $\ell_3$. The metric of a rotating black D3 brane is given by \cite{Russo:1998by}
\begin{align}
&ds^2=f_0^{-1/2}(-h_0dt^2+d\textbf{x}^2)+f_0^{1/2}\Biggl\{{\Delta dr^2\over\prod_{i=1}^3\left(1+{\ell_i^2\over r^2}\right)-{2m\over r^4}}+\\\nn
&+r^2\left[\Delta_1 d\q^2+\Delta_2\cos\q^2d\psi^2-2{\ell_2^2-\ell_3^2\over r^2}\cos\q\sin\q\cos\psi\sin\psi d\q d\psi+\right.\\\nn
&\left.+\left(1+{\ell_1^2\over r^2}\right)\sin\q^2d \phi_1^2+\left(1+{\ell_2^2\over r^2}\right)\cos\q^2\sin^2\psi d\phi_2^2+\left(1+{\ell_3^2\over r^2}\right)\cos^2\q\cos^2\psi d\phi_3^2+\right.\\\nn
&\left.+{2m\over r^6\Delta f_0}(\ell_1\sin^2\q d\phi_1+\ell_2\cos^2\q\sin^2\psi d\phi_2+\ell_3\cos^2\q\cos^2\psi d\phi_3)^2\right]+\\\nn
&-{4m\cosh \a\over r^4\Delta f_0}dt(\ell_1\sin^2\q d\phi_1+\ell_2\cos^2\q\sin^2\psi d\phi_2+\ell_3\cos^2\q\cos^2\psi d\phi_3)\Biggl\}
\end{align}
where $d\textbf{x}^2$ denotes the Euclidean metric of the longitudinal ${\bf R}^3$ and
\begin{align}
&h_0=1-{2m\over r^4\Delta}\quad,\quad f_0=1+{2m\sinh^2\a\over r^4\Delta},\\\nn
&\Delta=1+{\ell_1^2\over r^2}\cos^2\q+{\ell_2^2\over r^2}(\sin^2\q\sin^2\psi+\cos^2\psi)+{\ell_3^2\over r^2}(\sin^2\q \cos^2\psi+\sin^2\psi)+\\\nn
&+{\ell_2^2\ell_3^2\over r^4}\sin^2\q+{\ell_1^2\ell_3^2\over r^4}\cos^2\q\sin^2\psi+{\ell_1^2\ell_2^2\over r^4}\cos^2\q\cos^2\psi,\\\nn
&\Delta_1=1+{\ell_1^2\over r^2}\cos^2\q+{\ell_2^2\over r^2}\sin^2\q \sin^2\psi+{\ell_3^2\over r^2}\sin^2\q\cos^2\psi,\\\nn
&\Delta_2=1+{\ell_2^2\over r^2}\cos^2\psi+{\ell_3^2\over r^2}\sin^2\psi
\end{align}
The parameters $\a$ and $m$ are related to the D3-brane charge $N$ by
\be
4\pi g_s N\a'^2=2m\cosh\a \sinh\a.
\ee
The horizon $r=r_H$ is given by the largest real root of
\be\label{rotd3horizons}
\prod_{i=1}^3(r^2+\ell_i^2)-2mr^2=0.
\ee
The extremal configuration is reached if we take simultaneously $\a\rightarrow\infty$ and $m\rightarrow0$, keeping fixed  $ L^4=2m \sinh^2\a$. If $\ell_i\neq 0$, the limit $m\rightarrow 0$ exposes a naked singularity \cite{Russo:1998by}. 

The null Hamiltonian in the extremal case reads
\begin{align}
\mathcal{H}&=-f_0^{1/2}E^2+f_0^{-1/2}\Biggl\{{\prod_{i=1}^3\r_i^2\over r^6\Delta}P_r^2+{J_1^2\over \r_1^2 \sin^2\q}+{J_2^2\over \r_2^2 \cos^2\q \sin^2\psi}+{J_3^2\over \r_3^2\cos^2\q \cos^2\psi} +\\\nn
& +{r^2(\Delta_2  P_\q^2 +   \Delta_1 \sec^2\q P_\psi^2) + 2(\ell_2^2-\ell_3^2)\sin\psi\cos\psi \tan\q P_\q P_\psi \over r^4\Delta_1 \Delta_2-(\ell_2^2-\ell_3^2)^2\cos^2\psi\sin^2\q\sin^2\psi}\Biggl\}
\end{align}
where $\r_i^2=r^2+\ell_i^2$. In general it is not separable.

If we restrict attention on geodesics in the $\q=\psi=0$ plane, where $\partial \mathcal{H}/ \partial \q=\partial \mathcal{H}/ \partial \psi=0$ so that one can consistently set $P_\q=P_\psi=0$, the metric drastically simplifies
\begin{align}
&ds^2=-f_0^{-1/2}dt^2+f_0^{1/2}\Delta_3^{-1}dr^2+r^2\Delta_3f_0^{1/2}d\phi_3^2\\\nn
&\Delta_3=1+{\ell_3^2\over r^2}\quad,\quad \Delta=\Delta_1\Delta_2\quad,\quad h_0=1\quad,\quad f_0=1+{L^4\over \Delta r^4}
\end{align}
In Hamiltonian form, the system is separable and the radial momentum can be expressed in terms of the radial variable $r$ and the conserved quantities i.e. the energy $E$ and the relevant angular momentum $J=J_3$, since $J_1=J_2=0$,  
\be
{r^4 \Delta_3^2 P_r^2\over E^2}=r^4\Delta_3 f_0-b^2r^2=r^4\Delta+{L^4\over \Delta}-b^2 r^2
\ee
where $b={J/E}$.
If we take for simplicity $\ell_1=\ell_2=\ell_3=\ell\le L$, so that $\Delta_1=\Delta_2=\Delta_3$, we find
\be\label{zerospr}
P_r^2={E^2\over r^4 \left(1+{\ell^2\over r^2}\right)^3}\Bigg[(r^2+\ell^2)^2-b^2(r^2+\ell^2)+L^4\Bigg]
\ee
The radial action, in the coordinate $z= r^2+\ell^2$ reads:
\be
S_r=\int P_r dr=\int{E\over 2}{dz\over z^{3/2}}\sqrt{z^2-b^2 z+L^4}
\ee
This integral is invariant under the transformation $z\rightarrow L^4/ z$, or in the original radial coordinate
\be
r^2+\ell^2\rightarrow {L^4\over r^2+\ell^2}
\ee
The critical impact parameter and the critical radius (that determines the photon-sphere in this plane) are the critical points of $P_r$, thus looking at \eqref{zerospr}:
%\be
%r^4\left(1+{l^2\over r^2}\right)^2+L^4-b^2 r^2\left(1+{l^2\over r^2}\right)=0
%\ee
%whose solutions are:
\be
r_c^2+\ell^2={1\over2}\Bigg[ b_c^2\pm\sqrt{b_c^4-4L^4}\Bigg]
\ee
So the critical impact parameter is the same as in the non-rotating case
\be
b_c=\sqrt{2} L\quad,\quad r_c^2=L^2-\ell^2
\ee
The significance of the CT symmetry in the present context is not totally clear since the range of validity is limited to $\ell^2 r^2 < L^4- \ell^4$ that, reassuringly,  includes the photon-sphere for $L>\ell$.

\section{No horizon, No CT symmetry, notwithstanding the photon-sphere!}\label{RotD3D3fuzz}

As shown in \cite{Bianchi:2021yqs} and above, CT transformations exchange null infinity with the horizon keeping the photon-sphere fixed. Thus we should not expect smooth horizonless geometries like JMaRT \cite{Jejjala:2005yu} (and their supersymmetric limit GMS \cite{Giusto:2004id, Giusto:2004ip}) nor (circular) fuzz-balls in $D=6$ to enjoy CT symmetry despite the presence of a photon-sphere. One may try to identify a generalisation of CT transformations that exchange infinity with the cap but the different nature of the two regions presagite the failure of the attempt, as we will show in the following. In order to illustrate the problem, we briefly analyse massless geodesics    in the smooth horizonless backgrounds of the JMaRT (and GMS) family and in the circular fuzz-balls in $D=6$.

\subsection{JMaRT and GMS limit in $d=6$} 
JMaRT solutions \cite{Jejjala:2005yu} are smooth horizonless geometries of 3-charge systems. They are `over-rotating' to be considered {\it bona fide} micro-states of BHs. They are known to undergo an ergo-region instability \cite{Cardoso:2005gj, Maggio:2018ivz} and a Penrose process \cite{Bianchi:2019lmi} that should produce a BPS supersymmetric GMS solution \cite{Giusto:2004id, Giusto:2004ip} as stable remnant \cite{Chakrabarty:2019ujg}. 
The six-dimensional metric of JMaRT can be written as:
\be
ds_6^2=ds_5^2+\left(dy-{A\over\sqrt{3}}\right)^2
\ee
where in the extremal limit and in the equatorial plane ($\q=\pi/2$) 
\be
ds_5^2=-Z^{-2}(dt+\a)^2+Zds_4^2
\ee
%with
\begin{align}
&A={Q\over Z f}\left[-dt+{\sqrt{Q}\over 2}\left({a_\psi\over a_\phi}+{a_\phi\over a_\psi}+{1\over Q}a_\phi a_\psi\right)d\phi\right] \quad,\quad Z=1+{Q\over f} \\\nn
&f=r^2-a_\psi^2  \quad,\quad \alpha={\sqrt{Q}\over 2f}\left[Q\left({a_{\psi}\over a_{\phi}}+{a_{\phi}\over a_{\psi}}\right)+(1+2Z)a_{\psi}a_{\phi}\right]d\phi
\end{align}
%and 
\be
ds_{4,extr}^2={f dr^2\over r^2+a_{\phi}^2-a_{\psi}^2}+\left({a_{\phi}^2a_{\psi}^2\over f}+r^2+a_{\phi}^2\right)d\phi^2
\ee The components of the metric are easily identified
\begin{align}
&g_{tt}=-Z^{-2}+{Q^2\over 3Z^2 f^2}\quad,\quad g_{yy}=1\quad,\quad g_{rr}={Z f\over r^2+a_\phi^2-a^2_\psi}\\\nn
&g_{\phi\phi}={Q^3\over12Z^2 f^2}\left({a_\psi\over\a_\phi}+{a_\phi\over\a_\psi}+{a_\phi a_\psi\over Q}\right)^2-{Z^{-2}Q\over 4 f^2}\left[Q\left({a_\psi\over\a_\phi}+{a_\phi\over\a_\psi}\right)+(1+2Z)a_\psi a_\phi\right]^2 +\\\nn
&\quad\quad+Z\left({a_\phi^2a_\psi^2\over f}+r^2+a_\phi^2\right)\\\nn
&g_{t\phi}=g_{\phi t}=-{Q^{5/2}\over 6 Z^2 f^2}\left({a_\psi\over\a_\phi}+{a_\phi\over\a_\psi}+{a_\phi a_\psi\over Q}\right)-{Z^{-2}\sqrt{Q}\over2 f}\left[Q\left({a_\psi\over\a_\phi}+{a_\phi\over\a_\psi}\right)+(1+2Z)a_\psi a_\phi\right]\\\nn
&g_{ty}=g_{yt}={Q\over\sqrt{3} Z f}\quad,\quad g_{\phi y}=g_{y\phi}=-{Q^{3/2}\over2\sqrt{3}Zf}\left({a_\psi\over\a_\phi}+{a_\phi\over\a_\psi}+{a_\phi a_\psi\over Q}\right)
\end{align}
Setting for simplicity $P_y=0$, the relevant element of the inverse metric are also easy to determine
%\be 
%g g^{tt}=g_{rr}(g_{\phi\phi}-g^2_{y\phi})\quad,\quad g g^{t\phi}=g g^{\phi t}=g_{rr}(-g_{t\phi} + g_{ty} g_{y\phi})
%\ee
%$$
%gg_{rr}=-g_{t\phi}^2 + 2g_{ty}g_{t\phi} g_{y\phi} - g_{tt} g_{y\phi}^2+ g_{tt} g_{\phi\phi} -g_{ty}^2 g_{\phi\phi}$$
%$$gg_{\phi\phi}=g_{rr} (g_{tt}- g_{ty}^2)$$
%$$ g=-g_{rr} g_{t\phi}^2+2g_{rr}g_{ty}g_{t\phi}g_{y\phi} -g_{rr}g_{tt} g_{y\phi}^2+ g_{rr}g_{tt} g_{\phi\phi} -g_{rr} g_{ty}^2g_{\phi\phi}
%$$
and the Hamiltonian for null geodesics reads:
\begin{align}
&P_r^2=g_{rr}(2E J g^{t\phi}-E^2 g^{tt}-J^2g^{\phi\phi})={E^2\left(\sum_{k=0}^3A_{2k} r^{2k}\right)\over4a_\f^2a_\psi^2 r^2(r^2-a_\psi^2+a_\f^2)^2}\\\nn
&A_6=4a_\f^2 a_\psi^2 \\\nn
&A_4=4a_\f^2a_\psi^2(a_\f^2-2a_\psi^2-b^2+3Q)\\\nn
&A_2=4a_\f a_\psi\Big[-a_\psi^2 bQ^{3/2}-a_\f^2 b\sqrt{Q}(3a_\psi^2+Q)-a_\f^3(a_\psi^3-3a_\psi Q)+a_\f(a_\psi^5+a_\psi^3(2b^2-3Q)+3a_\psi Q^2)\Big]\\\nn
&A_0=-\Big[2a_\f a_\psi^3 b-a_\psi^2 Q^{3/2}+a_\f^2\sqrt{Q}(-3a_\psi^2+Q)\Big]^2
\end{align}
So the asymptotics of the radial action at infinity and near the cap $r_0^2= a_\psi^2-a_\phi^2$ are very different
\begin{align}
&S_r\sim E\int^\infty dr \qquad {vs} \qquad S_r\sim E\beta^2\int_{r_0} d\log(r-r_0)\quad,\quad \\\nn
&\beta^2={2ba_\phi^3a_\psi+Q^{3/2}a_\psi^2-\sqrt{Q}a_\phi^2(Q+3a_\psi^2)\over4a_\phi a_\psi r_0^{2}}
\end{align}
which seems to exclude any conceivable CT symmetry acting by conformal inversions. The only chance would be taking 
\subsection{D1/D5 fuzz-ball}

With little hope, we now explore case of circular fuzz-balls of 2-charge BHs and identify a CT-like transformation that gets close to being a symmetry, but we should recognise our inability that can be ascribed to the absence of the horizon that may reflect into new features in the GW ring-down signal such as echoes \cite{Cardoso:2017cqb, Guo:2017jmi, Bena:2019azk, Mayerson:2020tpn, Ikeda:2021uvc, Bah:2021jno} and modified memories \cite{Strominger:2014pwa, Pasterski:2015tva, Strominger:2017zoo, Addazi:2020obs, Aldi:2020qfu, Aldi:2021zhh} or else in their multipoles \cite{Bena:2020see, Bianchi:2020bxa, Bena:2020uup, Bianchi:2020miz}.

The 6-d metric of a circular D1/D5 or, equivalently, D3-D3' fuzz-ball reads \cite{Lunin:2001jy, Mathur:2009hf, Bianchi:2016bgx, Bianchi:2017bxl}
\begin{align}
ds^2&=H^{-1}\left[-\left(dt+\omega_\phi d\phi\right)^2+\left(dz+\omega_\psi d\psi\right)^2\right]+\\\nonumber
&+H\left[\left(\rho^2+a^2 \cos^2\q\right)\left({d\rho^2\over\rho^2+a^2}\right)+\rho^2\cos^2\q d\psi^2+\left(\rho^2+a^2\right)\sin^2\q d\phi^2\right]
\end{align}
where $H=\sqrt{H_1H_5}$, with
\be
\label{fuzzmetric}
H_i=1+{L_i^2\over\rho^2+a^2\cos^2\q}\quad, \quad 
\omega_\phi={aL_1L_5\sin^2 \q\over\rho^2+a^2\cos^2\q}\quad,\quad \omega_\psi={aL_1L_5\cos^2\q\over\rho^2+a^2\cos^2\q}.
\ee
The metric \eqref{fuzzmetric} has no explicit dependence on $t$, $z$, $\phi$ and $\psi$. As a result, the conjugate momenta $P_t= -E$, $P_z=P$, $P_\phi = J_\phi$ and $P_\psi = J_\psi$ are conserved.

%The system, governed by $\mathcal{L}{1\over 2}{ds^2\over d\tau^2}$, admits the four isometries and as many invariants for the geodetic motion:
%\begin{align}
%-E&={\partial \mathcal{L}\over \partial \dot{t}}=-{1\over H}\left(\dot{t}+\omega_\phi \dot{\phi}\right)\\\nonumber
%P&={\partial \mathcal{L}\over \partial \dot{z}}={1\over H}\left(\dot{z}+\omega_\psi\dot{\psi}\right)\\\nonumber
%J_\phi&={\partial \mathcal{L}\over \partial \dot{\phi}}=-{\omega_\phi\over H}\left(\dot{t}+\omega_{\phi}\dot{\phi}\right)+H\left(\rho^2+a^2\right)\sin^2\q \dot{\phi}\\\nonumber
%J_\psi&={\partial \mathcal{L}\over \partial \dot{\psi}}={\omega_\psi\over H}\left(\dot{z}+\omega_{\psi}\dot{\psi}\right)+H\left(\rho^2+a^2\right)\cos^2\q \dot{\psi}.
%\end{align}
Quite remarkably, very much as for KN or STURBHs, the system is separable \cite{Bena:2016ypk, Bianchi:2017sds, Bianchi:2018kzy}. In order to expose this property, one introduces the conjugate momenta 
\be 
\label{canconjmom1}
P_\rho={\partial \mathcal{L}\over \partial \dot{\rho}}={H\left(\rho^2+a^2\cos^2\q\right)\dot{\rho}\over\rho^2+a^2}\quad , \quad 
P_\q={\partial \mathcal{L}\over \partial \dot{\q}}=H\left(\rho^2+a^2\cos^2\q\right)\dot{\q}.
\ee
and writes the (zero) mass-shell condition in Hamiltonian form ${\cal H}= g^{\mu\nu} P_\mu P_\nu = 0$ 
% reasonable form:
%\begin{align}
%&\left(\rho^2+a^2\right)P_\rho^2-E^2\left(\rho^2+L_1^2+L_5^2+{L_1^2L_5^2\over\rho^2+a^2}\right)+P^2\left(\rho^2+L_1^2+L_5^2+{L_1^2L_5^2\over \rho^2}\right)+\\\nonumber
%&+{2EJ_\phi aL_1L_5\over \rho^2+a^2}-{2PJ_\psi a L_1L_5 \over \rho^2}-{a^2J_\phi^2\over\rho^2+a^2}+{a^2 J_\psi^2\over \rho^2}+P_\q^2-E^2a^2\cos^2\q+P^2a^2\cos^2\q+\\\nonumber
%&+{J^2_\phi\over \sin^2\q}+{J_\psi^2\over \cos^2\q}=0
%\end{align}
that can be separated into two equations
\begin{align}\label{fuzzeq}
K^2&=P_\q^2-\left(E^2-P^2\right)a^2\cos^2\q+{J^2_\phi\over \sin^2\q}+{J_\psi^2\over \cos^2\q}\\\nonumber
-K^2&=\left(\rho^2+a^2\right)P_\rho^2+{\left(P L_1L_5-aJ_\psi\right)^2\over\rho^2}-{\left(E L_1 L_5-aJ_\phi\right)^2\over\rho^2+a^2}-\left(\rho^2+L_1^2+L_5^2\right)\left(E^2-P^2\right)
\end{align}
where $K^2$ is the separation constant.
For brevity we will set
\be
\mathcal{E}^2=E^2-P^2\quad,\quad B^2_{\psi}={(P L_1 L_5-a J_\psi)^2\over\mathcal{E}^2a^2}\quad,\quad B^2_\f={(E L_1 L_5-a J_\f)^2\over\mathcal{E}^2a^2 }\quad,\quad b^2={K^2\over \mathcal{E}^2}
\ee
By the second (radial) equation in \eqref{fuzzeq} the radial action is:
\be
S_\r=\int d\r P_\r=\int{\mathcal{E}d\r\over\sqrt{\r^2+a^2}}\sqrt{{a^2B^2_\f\over\r^2+a^2}-{a^2B^2_\psi\over\r^2}+\r^2+L_1^2+L_5^2-b^2}
\ee
In terms of the variable $\r^2=x$ and if $B^2_\psi=0$ it becomes:
\be
\int{\mathcal{E}dx\over2\sqrt{x}(x+a^2)}\sqrt{x^2+x(L_1^2+L_5^2-b^2+a^2)+a^2B^2_\f+a^2L_1^2+a^2L_5^2-a^2b^2}
\ee
One can envisage a generalised CT inversion of the form \be
x+a^2\rightarrow {x_0+a^2\over x+a^2}\ee
with some $x_0$, alas this is not leave the radial action invariant for any choice of $x_0$, except for the case $a=0$ that produces a 2-charge small BH.

\section{(A)dS KN vs HE and CT invariance}\label{AdSKNsec}

Let us pass on to consider AdS$_4$ KN BHs and study scalar wave propagation in this backgrounds. As shown in \cite{Bianchi:2021mft}, in order to have the minimal number (i.e. four) of regular singular points in the radial equation, it is convenient to choose a `tachyonic' scalar with mass $\mu^2L^2 = -2 $. The mass is above the BF bound $\mu_{BF}^2L^2=-9/4$. Scalar fields of this kind are dual to boundary conformal operators of dimension $\Delta=1,2$ such as  $\phi^2$ or $\psi^2$  \cite{Aharony:2008ug, Bianchi:2010mg}, and correspond to `internal' fluctuations of the metric in the case of maximal supergravity in $S^7$. We spell out the conditions under which the relevant Heun equation (HE) with four regular singular points, corresponding to `real' and `complex' horizons, admit CT symmetry, exchanging horizons in pairs. This leads to a SHE (special Heun equation) with restricted parameters (but no confluence of singularities). However, quite disappointingly, when we translate the parameters of the SHE into BH parameters we find a trivial configuration with zero mass. We then switch to the extremal case as well as to generalised CT transformation that rescale the wave function.
  
\subsection{Heun equation for CT invariant AdS KN BHs}
The AdS-KN metric in 4-d is given by 
\be
ds^2=-{\Delta_r[dt-a d\f(1-\c^2)]^2\over\a^2\r^2}+{\Delta_\c[a dt-a\f(a^2+r^2)]^2\over \a^2\r^2}+\r^2\left({dr^2\over\Delta_r}+{d\c^2\over \Delta_\c}\right)
\ee
where $\c=\cos\q$ and
\begin{align}
&\Delta_r=(r^2+a^2)\left(1+{r^2\over L^2}\right)-2M r+Q^2={1\over L^2}(r-r_+)(r-r_-)(r-\r)(r-\bar{\r})\\\nn
&\Delta_\c=(1-\c^2)\left(1-{a^2\c^2\over L^2}\right)\quad,\quad\r^2=r^2+a^2\c^2\quad,\quad\a=1-{a^2\over L^2}.
\end{align}
The four roots of $\Delta_r$ satisfy
\be\label{sumroot}
r_++r_-+\r+\bar{\r}=0
\ee
and the relations with the BH parameters $a$, $Q$ and $M$:
\begin{align}\label{quartic}
&r_+ r_-+|\r|^2-(r_++r_-)^2=a^2+L^2\\\nn
&(|\r|^2-r_+ r_-)(r_++r_-)=2 M L^2\\\nn
&r_+ r_- |\r|^2=L^2(a^2+Q^2)
\end{align}
that easily allow to express $M$, $Q$ and $a$ in terms of $r_{\pm}$, $\rho$ and $\bar\rho$ at fixed $L$, while the inverse relations require solving a cubic equation and are more involved. For large $L>>M, a, Q$ two roots are real
\be
r_{\pm} \approx M \pm \sqrt{M^2-a^2-Q^2} + {\cal O}(L^{-2}) 
\ee
and two are complex conjugate
\be
\rho,\bar\rho = \pm i L -M \pm i {\cal O}(L^{-1})
\ee
 
The wave equation for a scalar can be separated. Setting
\be
\Phi(t,r,\c,\f) = e^{-i\omega t+im_\f\f}{\psi(r)S(\c)\over\sqrt{\Delta_r\Delta_\c}}
\ee
brings both radial and angular equations in canonical form.
The eigenvalue $K^2= \ell(\ell+1) + {\cal O}(a^2\omega^2)$ of the spheroidal harmonics equation for the angular part plays the role of a generalised angular momentum and enters the radial equation as a separation constant 
\begin{align}\label{AdSKNradial}
&\psi''(r) + Q_r(r) \psi(r)=0\\\nn
&Q_r(r)={\left(\hat{\omega}(a^2+r^2)-a \hat{m}\right)^2\over\Delta_r^2}-{K^2+{\mu^2r^2}\over\Delta_r}-{1\over2}{\Delta''_r\over \Delta_r}+{1\over 4}\left({\Delta'_r\over \Delta_r}\right)^2\\\nn
\end{align}
with $\hat{\omega}=\a \omega$, $\hat{m}=\a m_\f$.
For $\mu^2L^2=-2$ \eqref{AdSKNradial} has four regular singularities where $\Delta_r = 0$. For different choices of $\mu^2$ -- including 0 -- the resulting wave equation has more than 4 regular singular points. Performing the coordinate transformation
\be\label{maphe}
z={(r-r_+)(\r-r_-)\over (r-r_-)(\r-r_+)}
\ee
brings \eqref{AdSKNradial} in the standard form 
\be
\label{HeunEq}
\Biggl[\partial_z^2+\left({\gamma\over z}+{\delta\over z-1}+{\varepsilon \over z-t}\right)\partial_z+{\a \b z-p\over z(z-1)(z-t)}\Biggl]w(z)=0
\ee
which has four regular singular points at 0, 1, $t$ and $\infty$ that correspond to $r_+$, $\rho$, $\bar\rho$ and $r_-$ respectively. 
 
Since the curvature singularity at $r=0$ and the boundary $r=\infty$ are regular points of the radial equation, we should look for generalised CT transformations exchanging $r_+\leftrightarrow r_-$ and $\r\leftrightarrow\bar{\r}$ for instance. In the standard form we should impose invariance under $z\rightarrow t/z$  which requires
\be
\gamma= 1-\delta \quad, \quad \epsilon=\delta \quad, \quad \alpha\beta= 0
\ee 
and arbitrary $q$ and 
\be
t={(\bar\rho-r_+)(\r-r_-)\over (\bar\rho-r_-)(\r-r_+)}  
\ee
\eqref{HeunEq} can be put in canonical form if we introduce:
\be
w(z)=P(z) \psi(z)\quad\text{with}\quad P(z)=z^{-\gamma/2}(z-1)^{-\delta/2}(z-t)^{-\varepsilon/2}
\ee
and taking into account the restrictions on the parameters, it becomes:
\be\label{he2}
\psi''(z)+\Biggl[\frac{1-\delta ^2}{4z^2}+\frac{2 \delta -\delta ^2}{4 (z-1)^2}+\frac{2 \delta -\delta ^2}{4 (z-t)^2}+\frac{\delta ^2-\delta +2 p+\delta ^2 t-\delta  t +(2 \delta-\delta ^2 ) z}{2 (z-1) z (t-z)}\Biggl]\psi(z)=0
\ee
Inverting \eqref{maphe} and comparing \eqref{he2} with the radial equation for a scalar with mass $\mu^2L^2=-2$ in the AdS-KN background put severe constraints on the BH parameters as well as on $\omega$ and $m_\phi$. We derive the dictionary in Appendix \ref{appc}, by exploiting the recently established correspondence between BH perturbation theory  and quantum Seiberg-Witten curves \cite{Aminov:2020yma, Bianchi:2021xpr, Bonelli:2021uvf, Bianchi:2021mft, Bonelli:2022ten}. In the present case of AdS4-KN BHs, the relevant ${\cal N}=2$ SYM theory has gauge group $SU(2)$ and $N_f=4$ flavours of fundamental hypermultiplets with masses $m_f$. Moreover, imposing  $z\rightarrow t/z$ invariance (\`a la CT) requires $m_0=m_\infty$ and  $m_1=m_t$. 

Although beyond the scope of the present investigation, let us stress here that the analysis of QNMs and other observables associated to fluctuations around a variety of BHs, D-branes and fuzz-balls with flat or AdS asymptotics governed by the standard Heun Equation (HE) of various kinds, can be successfully tackled within this approach. Without dwelling into the details, let us pause and simply list, for the curious reader, the various interesting cases for the radial equation\footnote{Spheroidal harmonics in $D=4$ ($S^2$ sphere) and $D=4$ ($S^3$ sphere) are related to CHE and RCHE respectively. Nothing special happens to these angular equations in the extremal limit.} following\footnote{For convenience we split the flavours into $N_f=(N_L,N_R)$ as they appear in the Hanany-Witten setup of the ${\cal N}=2$ SYM theory \cite{Hanany:1996ie}.}  \cite{Aminov:2020yma, Bianchi:2021xpr, Bonelli:2021uvf, Bianchi:2021mft, Bonelli:2022ten}
\begin{itemize} 
\item Heun Eq (HE) with 4 regular singular points $\sim$  $N_f = 4 = (2_L,2_R)$ SW $\sim$ AdS$_4$ KN with $\mu^2L^2= -2$ (NO CT symmetry, in general)
\item Confluent Heun Eq (CHE) with 2 regular and 1 irregular singular points $\sim$  $N_f = 3 = (2_L,1_R)$ SW $\sim$ flat KN or STURBH (NO CTsymmetry, in general)
\item Reduced Confluent Heun Eq (RCHE) with 2 regular and 1 irregular singular points $\sim$  $N_f = 2 = (2_L,0)$ SW $\sim$ CCLP BHs and D1/D5 or D3/D3' circular fuzz-balls  (NO CT simmetry, in general)
\item Doubly Confluent Heun Eq (DCHE) with 2 irregular singular points $\sim$  $N_f = 2 = (1_L,1_R)$ SW $\sim$ Extremal KN BHs and ESTURBHs (CT symmetry)
\item Reduced Doubly Confluent Heun Eq (RDCHE) with 2 irregular singular points $\sim$  $N_f = 1 = (1_L,0)$ SW $\sim$ Extremal CCLP or BMPV BHs  (CT symmetry)
\item Doubly Reduced Doubly Confluent Heun Eq (DRDCHE) with 2 irregular singular points $\sim$  $N_f = 0 = (0_L,0_R)$ SW $\sim$ Extremal D3-branes, D1/D5 or D3/D3' BHs, and D3-D3-D3-D3 BHs  (CT symmetry)
\end{itemize}

Going back to our problem, in the end it turns out that only trivial values for the BH parameters (as for example $M=0$) are compatible with CT symmetry. 

We then have two options: consider extremal KN BH (in subsection \ref{eadshnsec}), or gain some flexibility by performing a Weyl transformation of HE so as to introduce new parameters in the equation (in subsection \ref{madsHe}).

\subsection{Extremal AdS KN vs restricted confluent Heun equation}\label{eadshnsec}

Some simplifications occur for extremal BHs with $r_+=r_- = r_H$ still with $\rho\neq \bar\rho$ since we consider $L>>M$. For $\rho=\bar\rho$ one would get a doubly confluent HE (DCHE) that can be easily shown to admit CT symmetry in general. The extremal mass is \cite{Caldarelli:1999xj}:
\be\label{adsknextr}
 M_{\rm extr}={L\over 3\sqrt{6}}\left(\sqrt{\left(1{+}{a^2\over L^2}\right)^2{+}{12\over L^2}(a^2{+}Q^2)}{+}{2a^2\over L^2}{+}2\right)\left(\sqrt{\left(1{+}{a^2\over L^2}\right)^2{+}{12\over L^2}(a^2{+}Q^2)}{-}{a^2\over L^2}{-}1\right)^{1\over2}.
 \ee
The relevant equation turns out to be the CHE (confluent Heun equation) that reads:
\be
\label{ConflHeunEq}
\Bigg[\partial_z^2+\left({\gamma\over z}+{\delta\over z-1}+\e\right)\partial_z+{\a z-q\over z(z-1)}\Bigg]W(z)=0.
\ee
If we define $W(z)=e^{-{\e z\over2}}F(z)$ and impose symmetry under $z\rightarrow1-z$, the parameters of the equation must obey to $\gamma=\d$ and $\a=\gamma \e$. So the equation becomes:
%\be
%\Bigg[\partial_z^2+\left({\gamma\over z}+{\delta\over z-1}\right)\partial_z+\left(-{\e^2\over 4}+{2q-\gamma \e\over 2z}+{2\a-2q-\d \e\over 2(z-1)}\right)\Bigg]F(z)=0
%\ee
%If we impose the symmetry under , , so that the equation appears:
\be
\Bigg[\partial_z^2+\gamma\left({1\over z}+{1\over z-1}\right)\partial_z+\left({-{\e^2\over 4}+{\tilde{q}\over z(z-1)}}\right)\Bigg]F(z)=0\quad,\quad \tilde{q}={\gamma \e\over 2}-q.
\ee
In order to establish a dictionary with extremal AdS KN BH, we perform the mapping:
\be\label{rcoor}
z={(r-\r)(\bar{\r}-r_H)\over (r-r_H)(\bar{\r}-\r)}
\ee
%so that the equation becomes:
%\be
%\Bigg[\partial_r^2+\left({2(1-\gamma)\over r-r_H}+{\gamma\over r-\r}+{\gamma\over r-\bar{\r}}\right)\partial_r+{\tilde{q}(\r-r_H)(\bar{\r}-r_H)\over(r-\r)(r-\bar{\r})(r-r_H)^2}-{\e^2(\r-r_H)^2(\bar{\r}-r_H)^2\over4(r-r_H)^4(\r-\bar{\r})^2}\Bigg]F(r)=0
%\ee
%If we define $F(r)=A(r) F(r)$ with:
%\be
%A(r)=(r-r_H)^{\gamma-1}(r-\r)^{-{\gamma\over2}}(r-\bar{\r})^{-{\gamma\over2}}.
%\ee
In canonical form the radial equation reads
\begin{align}\label{QrH}
&\psi''(r)+Q_r(r)\psi(r)=0\\\nn
&Q_r(r)={\tilde{q}+\g-\g^2\over(r-r_H)^2}+{2\g-\g^2\over 4(r-\r)^2}+{2\g-\g^2\over 4(r-\bar{\r})^2}-{\e^2(3\r+\bar{\r})^2(\r+3\bar{\r})^2\over64(\r-\bar{\r})^2(r-r_H)^4}+{8(\tilde{q}+\g-\g^2)(\r+\bar{\r})\over(3\r+\bar{\r})(\r+3\bar{\r})(r-r_H)}+\\\nn
&+{(\g-4)\g\r+(4-5\g)\g\bar{\r}+2\tilde{q}(\r+3\bar{\r})\over 2(\r-\bar{\r})(3\r+\bar{\r})(r-\r)}+{(\g-4)\g\bar{\r}+(4-5\g)\g\r+2\tilde{q}(3\r+\bar{\r})\over2(\bar{\r}-\r)(\r+3\bar{\r})(r-\bar{\r})}
\end{align}
In terms of radial coordinates, the exchange symmetry $z\rightarrow 1-z$ translates into:
\be
r-r_H\rightarrow {(r-r_H)(r_H-\r)(r_H-\bar{\r})\over 2(\r+\bar{\r})r+r_H^2-\r\bar{\r}}
\ee
We start the comparison with \eqref{AdSKNradial} by studying the residues of the double poles in $\r$ and $\bar{\r}$.
%\begin{align}
%&\lim_{r\rightarrow \r}\Big[(r-\r)^2Q_r\Big]={L^4\Big[\hat{\omega}(\r^2+a^2)-a\hat{m}\Big]^2\over(\r-r_H)^4(\r-\bar{\r})^2}+{1\over 4}={2\g-\g^2\over4}\\\nn
%&\lim_{r\rightarrow \bar{\r}}\Big[(r-\bar{\r})^2Q_r\Big]={L^4\Big[\hat{\omega}(\bar{\r}^2+a^2)-a\hat{m}\Big]^2\over(\bar{\r}-r_H)^4(\bar{\r}-\r)^2}+{1\over 4}={2\g-\g^2\over4}
%\end{align}
Imposing equality of the two residues we obtain
%a relation between the roots and the impact parameter $b_\f=m/\omega$:
\be
a^2-a b_\f={1\over 2}\Bigg[\r\bar{\r}+{(\r+\bar{\r})^2\over4}\Bigg]={1\over2}\left(a^2+L^2+4r_H^2\right)\quad,\quad(\g-1)^2=-{L^4\hat{\omega}^2\over(\r-\bar{\r})^2}.
\ee
%furthermore we obtain an expression for $\g$:
%\be
%(\g-1)^2=-{L^4\hat{\omega}^2\over(\r-\bar{\r})^2}
%\ee
By comparing the residues of the order four poles in $r_H$, we obtain an expression for $\e$
%\be
%\lim_{r\rightarrow r_H}\Big[(r-r_H)^4Q_r(r)\Big]={L^4\Big[\hat{\omega}(r_H^2+a^2)-a\hat{m}\Big]^2\over(r_H-\r)^2(r_H-\bar{\r})^2}=-{\e^2(3\r+\bar{\r})^2(\r+3\bar{\r})^2\over64(\r-\bar{\r})^2}
%\ee
%so we 
\be
\e={\pm4 i\hat{\omega}L^2(\r-\bar{\r})\over(3\r+\bar{\r})(3\bar{\r}+\r)}
\ee
If we compare the residues of the double and the single poles in $r_H$, we obtain:
\be
%{1\over2}\lim_{r\rightarrow r_H}{d^2\over dr^2}\Bigg[(r-r_H)^4Q_r(r)\Bigg]=
2{L^4\hat{\omega}^2-2K^2L^2+(\r+\bar{\r})^2\over(3\r+\bar{\r})(3\bar{\r}+\r)}=\tilde{q}+\g-\g^2
\ee
%\be
%{1\over6}\lim_{r\rightarrow r_H}{d^3\over dr^3}\Bigg[(r-r_H)^4Q_r(r)\Bigg]={16(\r+\bar{\r})\Big[-2K^2L^2+L^4\hat{\omega}^2+(\r+\bar{\r})^2\Big]\over(3\r+\bar{\r})^2(3\bar{\r}+\r)^2}={8(\tilde{q}+\g-\g^2)(\r+\bar{\r})\over(3\r+\bar{\r})(3\bar{\r}+\r)}
%\ee
The simple pole in $\r$:
\be
-{8K^2L^2(\r-\bar{\r})^2+(\r-\bar{\r})^3(5\r+3\bar{\r})+L^4(5\r-\bar{\r})(\r+3\bar{\r})\omega^2\over2(\r-\bar{\r})^3(3\r+\bar{\r})^2}={(\g-4)\g\r+(4-5\g)\g\bar{\r}+2\tilde{q}(\r+3\bar{\r})\over 2(\r-\bar{\r})(3\r+\bar{\r})}
\ee
%\be
%{4K^2L^2-8\r^2\over(\bar{\r}-\r)(3\r+\bar{\r})^2}+{3\bar{\r}-7\r\over2(\r-\bar{\r})(3\r+\bar{\r})}-{L^4(5\r^2+14\r\bar{\r}-3\bar{\r}^2)\omega^2\over2(\r-\bar{\r})^3(3\r+\bar{\r})^2}={(\g-4)\g\r+(4-5\g)\g\bar{\r}+2q(\r+3\bar{\r})\over 2(\r-\bar{\r})(3\r+\bar{\r})}
%\ee
The relation that holds for $\bar{\r}$ is the same as the previous one after exchanging $\r$ and $\bar{\r}$.
%\be
%-{8K^2L^2(\r-\bar{\r})^2+(\bar{\r}-\r)^3(5\bar{\r}+3\r)+L^4(5\bar{\r}-\r)(\bar{\r}+3\r)\omega^2\over2(\bar{\r}-\r)^3(3\bar{\r}+\r)^2}={(\g-4)\g\bar{\r}+(4-5\g)\g\r+2\tilde{q}(3\r+\bar{\r})\over2(\bar{\r}-\r)(\r+3\bar{\r})}
%\ee
%\be
%{4K^2L^2-8\bar{\r}^2\over(\r-\bar{\r})(3\bar{\r}+\r)^2}+{3\r-7\bar{\r}\over2(\bar{\r}-\r)(\r+3\bar{\r})}+{L^4(5\bar{\r}^2+14\r\bar{\r}-3\r^2)\omega^2\over2(\r-\bar{\r})^3(\r+3\bar{\r})^2}={(\g-4)\g\bar{\r}+(4-5\g)\g\r+2q(3\r+\bar{\r})\over2(\bar{\r}-\r)(\r+3\bar{\r})}
%\ee
As a result, $\tilde{q}$ can be written (up to $\r\leftrightarrow \bar{\r}$ exchange) as:
\be
\tilde{q}=\pm{i L^2\hat{\omega}\over\r-\bar{\r}}-{L^4\hat{\omega}^2\over(\r-\bar{\r})^2}+{2\Big[-2K^2L^2+(\r+\bar{\r})^2+L^4\hat{\omega}^2\Big]\over(3\r+\bar{\r})(3\bar{\r}+\r)}
\ee
Thus $\omega$ and K remain free parameters, while the mass is dictated by extremality \eqref{adsknextr}.

Notice that the critical radii in the equatorial plane are the extremal points of
\be
P_4(r) = [r^2 - a (\alpha_L b - a)]^2 - (\alpha_L b - a)^2 \Delta_r(r)
\ee
 so that 
 \be
 r_c=\pm r_H \quad\text{with}\quad \alpha_L b_c = a + {r_H^2 \over a} 
 \ee
 In the end CT invariance works as much as possible as in the asymptotically flat case in that it keeps the photon-sphere fixed ($r_c=r_H$ in the equatorial plane) and exchanges the two complex horizons. Recall that both the AdS boundary ($r\rightarrow \infty$) and the curvature singularity ($r=0$) are regular points of the wave equation instead!

\subsection{Transformed Heun cs AdS KN with rescalings}\label{madsHe}

So far we have implicitly assumed that CT transformations ($z\rightarrow t/z$ or $z\rightarrow 1-z$) preserved the wave-function in the standard HE or CHE form. In principle this is not necessary and the function may undergo some (Weyl) rescaling. In order to account for this `flexibility'  let us go back to HE in \eqref{HeunEq} and 
%\be
%W''(z) +\left({\g\over z}+{\d\over z-1}+{\e\over z-t}\right){W'(z)}+{\a \b z-p\over z(z-1)(z-t)}W(z)=0
%\ee
perform the transformation\footnote{We allow only rescalings that keep the symmetry between $\rho$ and $\bar\rho$ i.e. $z=1$ and $z=t$.} $W(z)=z^a(z-1)^b(z-t)^bf(z)$. One gets a new HE with parsmeters
%\begin{align}\label{modHe}
%&f''(z)+\left({2a+\g\over z}+{2b+\d\over z-1}+{2b+\e\over z-t}\right)f'(z)+\left({a^2-a+a\g\over z^2}+{b^2-b+b\d\over(z-1)^2}+{b^2-b+b\e\over(z-t)^2}+\right.\\\nn&\left.+{-2 a b-p-2abt-b\g-bt\g-at\d-a\e\over t z}+{-2ab-2b^2+p+2abt-\a\b-b\g+bt\g-a\d-b\d+at\d-b\e\over(t-1)(z-1)}+\right.\\\nn&\left.+{-2ab-p+2abt+2b^2t+t \a\b-b\g+bt\g+bt\d-a\e+at\e+bt\e\over(t-1)t(z-t)}\right)f(z)=0
%\end{align}
%Now let's apply the CT inversion $z={t\over u}$ on modified HE \eqref{modHe}:
%\begin{align}\label{modHe}
%f''(u)+\left({2-2a-4b-\g-\d-\e\over u}+{2b+\d\over u-t}+{2b+\e\over u-1}\right)f'(u)+
%\end{align}
%Il we perform $z=t/u$, the equation becomes:
%\begin{align}
%&0=f''(u)+\left({2-2a-4b-\g-\d-\e\over u}+{2b+\e\over u-1}+{2b+\d\over u-t}\right)f'(u)+\\\nn
%&+\Bigg[\frac{a^2+4 a b+\alpha  \beta +a \gamma +a \delta +a \epsilon -a+4 b^2+2 b \gamma +2 b \delta +2 b \epsilon -2 b}{u^2}+\frac{b^2+b \delta -b}{(u-t)^2}+\frac{b^2+b \epsilon -b}{(u-1)^2}+\\\nn
%&+\frac{-2 a b t+2 a b+\alpha \beta +a \delta -a \delta  t-2 b^2 t+4 b^2+b \gamma +3 b \delta -b \gamma  t-2 b \delta  t+2 b t+b \epsilon -2 b-p}{(t-1) t (u-t)}+\\\nn
%&+\frac{-2 a b t+2a b-a t \epsilon +a \epsilon -4 b^2 t+2 b^2+b \gamma -b \gamma  t-b \delta  t-3 b t \epsilon +2 b t+2 b \epsilon -2 b+p-\alpha  \beta  t}{(t-1)(u-1)}+\\\nn
%&+\frac{2 a b t+2 a b+\alpha  \beta +a \delta +a t \epsilon +4 b^2 t+4 b^2+b \gamma +3 b \delta +b \gamma  t+b \delta  t+3 b t \epsilon -2 bt+b \epsilon -2 b-p+\alpha  \beta  t}{t u}\Bigg]f(u)
%\end{align}
\begin{align}
&\g\rightarrow 2a+\g\quad,\quad \d\rightarrow 2b+\d\quad,\quad \e\rightarrow 2b+\e\\\nn
&p\rightarrow (2ab+b\g)(1+t)+a \d t+a\e+p\\\nn
&\a \b\rightarrow \a \b+2b(2a+b+\g)+(\d+\e)(a+b)
\end{align}
Imposing symmetry under $z\rightarrow t/z$ requires
\be
\e=\d\quad,\quad \g=1-2(a+b)-\d\quad,\quad \a \b=-2(a+b) \d
\ee
%Thus, using the previous constraint on the parameters of the equation and if we introduce the coordinates \eqref{rcoor}, the equation reads:
%\begin{align}
%&f''(r)+\Bigg[{-a(a+2b+\d)\over(r-r_+)^2}+{-a(a+2b+\d)\over(r-r_-)^2}+{b(-1+b+\d)\over (r-\r)^2}+{b(-1+b+\d)\over(r-\bar{\r})^2}\Bigg]f'(r)+\\\nn
%&+\Bigg[{1\over r-r_+}\left(\frac{2 a^2+4 a b+2 b (2 b+\delta -1)-p}{r_+-r_-}+\frac{a \delta -b (2 b+\delta -1)+p}{r_+-\bar{\rho}}+\frac{a
%   \delta -b (2 b+\delta -1)}{r_+-\rho }\right)+\\\nn
%&+{1\over r-r_-}\left(\frac{2 a^2+4 a b+2 b (2 b+\delta -1)-p}{r_--r_+}+\frac{a \delta -b (2 b+\delta -1)+p}{r_--\rho }+\frac{a \delta -b (2 b+\delta
%   -1)}{r_--\bar{\rho}}\right)+\\\nn
%&+{1\over r-\r}\left(\frac{-2 a \delta +2 b^2-p}{\rho -\bar{\rho}}+\frac{a \delta -b(2b+\d-1)+p}{\rho -r_-}+\frac{a \delta -b(2b+\d-1)}{\rho
%   -r_+}\right)+\\\nn
%&+{1\over r-\bar{\r}}\left(\frac{-2 a \delta +2 b^2-p}{\bar{\rho} -\r}+\frac{a \delta -b (2 b+\delta -1)+p}{\bar{\rho}-r_+}+\frac{a \delta-b (2 b+\delta -1)
%   }{\bar{\rho}-r_-}\right)\Bigg]f(r)=0
%\end{align}
In the coordinates \eqref{maphe} SHE can be put in canonical form by introducing:
\be
f(r)=[(r-r_+)(r-r_-)]^{-{1\over2}(1-2b-\d)}[(r-\r)(r-\bar{\r})]^{-{1\over2}(2b+\d)} \psi(r)
\ee
so that:
\be
\psi''(r)+\sum_{i=1}^4\Bigg[{A_{2,i}\over(r-r_i)^2}+{A_{1,i}\over(r-r_i)}\Bigg]\psi(r)=0\\\nn
\ee
with
\begin{align}\label{residues1}
&A_{2,+}=A_{2,-}=-{1\over 4}(-1+\s+\d)(1+\s+\d)\qquad,\qquad A_{2,\r}=A_{2,\bar{\r}}={1\over4}(2-\d)\d\\\nn
&A_{1,+}={2p(r_+{-}\r)(r_-{-}\bar{\r}){-}(\d{+}\s{-}1)[{-}r_-^2\d{+}r_-r_+(1{-}3\d{+}\s){+}\r\bar{\r}(1{-}\d{+}\s){+}r_+^2(2{+}\d{+}2\s)]\over2(r_--r_+)(r_+-\r)(r_+-\bar{\r})}\\\nn
%&A_{1,-}={-2p(r_+-\r)(r_--\bar{\r})+(\d+\s-1)(-r_+^2\d+r_-r_+(1-3\d+\s)+\r\bar{\r}(1-\d+\s)+r_-^2(2+\d+2\s))\over 2(r_--r_+)(r_--\r)(r_--\bar{\r})}\\\nn
&A_{1,\r}={1\over2}\Biggl[{2p+\d(\d+2\s)\over\bar{\r}-\r}+{\d(\d+\s-1)\over\r-r_+}+{2p+\d(\d+\s-1)\over\r-r_-}\Biggl]\\\nn
&A_{1,-}=A_{1,+}(r_+\leftrightarrow r_-,\r\leftrightarrow\bar{\r})\quad,\quad A_{1,\r}=A_{1,\bar{\r}}(r_+\leftrightarrow r_-,\r\leftrightarrow\bar{\r})
%&A_{1,\bar{\r}}={1\over2}\left({2p+\d(\d+2\s)\over\r-\bar{\r}}+{\d(\d+\s-1)\over\bar{\r}-r_-}+{2p+\d(\s+\d-1)\over\bar{\r}-r_+}\right)
\end{align}
where $\s=2(a+b)$. 
%On the other side, Radial equation for AdS KN BH in canonical form reads:
%\begin{align}\label{AdSKN}
%&\psi''(r)+Q_r(r)\psi(r)=0\\\nn
%&Q_r(r)=\sum_{i=1}^4\Bigg[{B_{2,i}\over(r-r_i)^2}+{B_{1,i}\over(r-r_i)}\Bigg]\\\nn
%&Q_r(r)={1\over \Delta_r^2}\Bigg[\left(\hat{\omega}(a_J^2+r^2)-a_J\hat{m}\right)^2-\Delta_r(K^2+r^2M_{\Phi}^2)-{1\over 2}\Delta_r\Delta_r''+{1\over 4}\Delta_r^{'2}\Bigg]
%\end{align}
%where
%\be
%\hat{\omega}=\a_L \omega\quad,\quad \hat{m}=\a_Lm_\f\quad,\quad \a_L=1-{a_J^2\over L^2}
%\ee
From the perspective of the AdS KN wave equation \eqref{AdSKNradial}, the residues of the double poles are:
\be\label{residues2}
B_{2,i}=\lim_{r\rightarrow r_i}[(r-r_i)^2 Q_r(r)]={L^4[\hat{\omega}(r_i^2+a_J^2)-a_J\hat{m}]^2\over\prod_{j\neq i}(r_j-r_i)^2}+{1\over4}
\ee
%&B_{2,+}=\lim_{r\rightarrow r_+}[(r-r_+)^2 Q_r(r)]={L^4[\hat{\omega}(r_+^2+a_J^2)-a_J\hat{m}]^2\over(r_+-r_-)^2(r_+-\r)^2(r_+-\bar{\r})^2}+{1\over4}\\\nn
%&B_{2,-}=\lim_{r\rightarrow r_-}[(r-r_-)^2 Q_r(r)]={L^4[\hat{\omega}(r_-^2+a_J^2)-a_J\hat{m}]^2\over(r_--r_+)^2(r_--\r)^2(r_--\bar{\r})^2}+{1\over4}\\\nn
%&B_{2,\r}=\lim_{r\rightarrow \r}[(r-\r)^2 Q_r(r)]={L^4[\hat{\omega}(\r^2+a_J^2)-a_J\hat{m}]^2\over(\r-r_+)^2(\r-r_-)^2(\r-\bar{\r})^2}+{1\over4}\\\nn
%&B_{2,\bar{\r}}=\lim_{r\rightarrow \bar{\r}}[(r-\bar{\r})^2 Q_r(r)]={L^4[\hat{\omega}(\bar{\r}^2+a_J^2)-a_J\hat{m}]^2\over(\bar{\r}-r_+)^2(\bar{\r}-r_-)^2(\bar{\r}-\r)^2}+{1\over4}\\\nn

Since from \eqref{residues1} $A_{2,+}=A_{2,-}$, we must require that $B_{2,+}=B_{2,-}$ and the same for the residues of the double poles at $\r$ and $\bar{\r}$:
\begin{align}\label{delta}
{r_-^2+a_J^2-b_\f a_J\over r_+^2+a_J^2-b_\f a_J}=\pm{(r_--\r)(r_--\bar{\r})\over(r_+-\r)(r_+-\bar{\r})}\\\nn
{\r^2+a_J^2-b_\f a_J\over \bar{\r}^2+a_J^2-b_\f a_J}=\pm{(\r-r_+)(\r-r_-)\over(\bar{\r}-r_+)(\bar{\r}-r_-)}
\end{align}
where $b_\f=\hat{m}/\hat{\omega}=m/\omega$. By comparison, from the previous relations we can write a constraint on $b_\f$ (or equivalently $\omega$):
\be\label{a}
a_J^2-b_\f a_J=-(\r+\bar{\r})r_++|\r|^2=-(r_++r_-)\r+r_+r_-={1\over2}(r_+r_++|\r|^2)
\ee
On the other hand, exploiting the first two relations in \eqref{residues1} and \eqref{residues2} and plugging \eqref{a}, we have:
\be\label{deltaomega1}
\d+\s=\pm{2iL^2\hat{\omega}\over r_+-r_-}\quad,\quad1-\d=\pm{2iL^2\hat{\omega}\over\r-\bar{\r}}
\ee
Eqs. \eqref{deltaomega1} can be used to re express $\d$:
\be\label{delta2}
\d={\r-\bar{\r}-\z\s(r_+-r_-)\over\r-\bar{\r}+\z(r_+-r_-)}\quad\text{with}\quad \z=\pm1
\ee
The residues of the simple poles are
%can be computed by:
%\be
%B_{1,i}=\lim_{r\rightarrow r_i}{d\over dr}[(r-r_i)^2Q_r(r)]
%\ee 
%and they can be drastically simplified by using the relations:
%\begin{align}\label{rootsum}
%&\hat{m}={\hat{\omega}\over a}\Big[a^2-{1\over2}\left(r_+r_-+|\r|^2\right)\Big]\\\nn
%&r_++r_-+\r+\bar{\r}=0
%\end{align}
%so that:
\begin{align}
&B_{1,+}={(r_+{-}r_-)^2[{-}2K^2L^2{+}r_-^2{-}r_+^2{+}(r_-{+}\r)(r_+{+}\r)]{+}L^4[r_-^2{-}3r_-r_+{-}r_+^2+(r_-{+}r_+)\r{+}\r^2]\hat{\omega}^2\over2(r_+-r_-)^3(r_+-\r)(2r_++r_-+\r)}\\\nn
%&B_{1,-}={(r_+{-}r_-)^2(2K^2L^2+r_-^2-r_+^2-(r_-+\r)(r_++\r)){+}L^4(r_-^2+3r_-r_+-r_+^2-(r_-+r_+)\r-\r^2)\hat{\omega}^2\over2(r_+-r_-)^3(r_--\r)(2r_-+r_++\r)}\\\nn
&B_{1,\r}={(r_-{+}r_+{+}2\r)^2[{-}2K^2L^2{+}r_-^2{-}\r^2{+}(r_+{+}\r)(r_+{+}r_-)]{+}L^4[r_-^2{+}r_-r_+{+}r_+^2{+}5(r_-{+}r_+)\r{+}3\r^2]\hat{\omega}^2\over2(r_--\r)(r_+-\r)(r_-+r_++2\r)^3}\\\nn
&B_{1,-}=B_{1,+}(r_+\leftrightarrow r_-)\qquad,\qquad B_{1,\bar{\r}}=B_{1,\r}(\r\leftrightarrow \bar{\r})
%&B_{1,\bar{\r}}={(r_-{+}r_+{+}2\bar{\r})^2({-}2K^2L^2{+}r_-^2{-}\bar{\r}^2{+}(r_+{+}r_-)(r_+{+}\bar{\r})){+}L^4(r_-^2{+}r_-r_+{+}r_+^2{+}5(r_-{+}r_+)\bar{\r}{+}3\bar{\r}^2)\hat{\omega}^2\over2(r_--\bar{\r})(r_+-\bar{\r})(r_-+r_++2\bar{\r})^3}
\end{align}
Now it is possible to compute the parameter $p$ in two different ways, i.e. $A_{1,+}(p)=B_{1,+}$, $A_{1,-}(p)=B_{1,-}$. The two expressions for $p$ must be equal so, if we use \eqref{delta2}, one finds
\be
(r_-+r_+)\Big[(r_+-r_-)^2(r_-+r_++2\r)^2(1+\s)^2+L^4(r_-+r_++(r_+-r_-)\z+2\r)^2\hat{\omega}^2\Big]=0
\ee
Excluding $r_-=-r_+$ that would imply zero mass BH and using \eqref{deltaomega1}, \eqref{delta2}, as well as \eqref{sumroot} the previous expression becomes :
\be
(r_--r_+)(\r-\bar\rho)^2(1+\s)^2=0
\ee
Excluding extremality $r_-=r_+=r_H$, that would require $M=M_{\rm extr}$ already considered before, and $\rho= \bar\rho$, that would require $L<M$, eventually requires $\sigma=-1$. As a result of \eqref{deltaomega1} $\hat{\omega}=0$ and $\delta=1$. So the last parameter to determine is:
\be
p={K^2L^2+r_+r_- + \rho\bar\rho \over (r_+-\rho)(r_- -\bar\rho)}
\ee
$K$ remains a free parameter, while there are no conditions on the roots or, in other words, there are not constraints on $M$, $a$ and $Q$.

Once again we get a reduced form of CT invariance, valid only for `static' waves with ${\omega}=0$ and $m_\phi=0$.

\section{Conclusions}
\label{concl}

We have extended our investigation on CT conformal inversions \cite{CouchTorr, Cvetic:2020kwf, Cvetic:2021lss} of BHs and D-branes \cite{Bianchi:2021yqs} in various directions. 

First we have analysed asymptotically flat rotating charged BHs in $D=4$, in particular (near) extremal rotating BHs in STU supergavity (n-eSTURBHs) \cite{Cvetic:1996xz}, and found invariance for special choices of the charges. 

Second we have studied scalar wave equations in these backgrounds and identified the near super-radiant modes a.k.a. zero damping modes in that display a very slow fall off at late times and thus represent a rather distinctive feature of the GW signal emitted in the mergers of near-extremal BHs. 

We have then considered rotating BHs in $D=5$ \cite{Chong:2004na, Chong:2005hr} and rotating D3-branes \cite{Russo:1998by} and found invariance under generalised CT inversions for special choices of the angular momenta. Not surprisingly we have not found any similar symmetry for smooth horizonless geometries such as JMaRT \cite{Jejjala:2005yu}, their supersymmetric GMS version \cite{Giusto:2004id, Giusto:2004ip} or circular fuzz balls \cite{Lunin:2001jy, Mathur:2009hf}. Notwithstanding the presence of photon-rings, the lack of a horizon and the very different behaviours at infinity and at the cap are the reasons behind the failure.

Finally we have considered scalar perturbations of KN BHs in AdS$_4$ and found that the conditions for CT invariance are too stringent for generic AdS KN BHs unless one allows for a rescaling of the wave function and focuses on `static' waves. In the extremal case \eqref{adsknextr}, $m_\phi$ is related to $\omega$ and $K$, very much as the `critical' impact parameters of a massless probe are related to the `critical' radii of the photon-rings. 

In all cases CT invariance keeps the photon-halo fixed and exchanges either the horizon and null infinity (for flat asymptotics) or two complex horizons (for KN BHs in AdS).

The relevant scalar wave equations can be separated and both the radial and angular parts, that generalise the celebrated Regge-Wheeler-Zerilli and Teukolsky equations, can be brought in the form of Heun equation or confluent versions thereof.  At various points, we have also exploited the surprising connection with quantum SW curves for ${\cal N}=2$ SYM with $SU(2)$ gauge group and $N_f$ fundamental hypermultiplets
\cite{Aminov:2020yma, Bianchi:2021xpr, Bonelli:2021uvf, Bianchi:2021mft, Bonelli:2022ten}.

Since for rotating objects generalised CT transformations depend on the impact parameter(s), the relation between scattering angle for geodesics outside the photon-sphere and in-spiralling angle for geodesics inside the photon-sphere includes a boundary term $\Delta\phi_{bdry}(E,J) = \Delta\phi_{fall}(E,J)-\Delta\phi_{scatt}(E,J)$, determined by $\partial r_c/\partial J \neq 0$. In general, the observables are encoded in the full action that involves both a radial and a non-trivial angular part, although the latter plays only a marginal role in CT inversions.
 
The dynamics of rotating BHs at higher orders in $G_N$ (Post-Minkowskian) or in $v/c$ (post-Newtonian) -- the two being related by virial theorem -- is under very active investigation \cite{Bini:2021gat, Bern:2021yeh, Jakobsen:2021zvh,  Kol:2021jjc, Cho:2021arx, Brandhuber:2021eyq, Bautista:2021wfy}. Classical gravity may be extracted from quantum amplitudes \cite{Bjerrum-Bohr:2018xdl, KoemansCollado:2018hss}. New soft theorems  \cite{He:2014laa, Cachazo:2014fwa, Strominger:2017zoo}, valid even in the string context \cite{Bianchi:2015yta, Bianchi:2015lnw}, can help understanding radiation reaction, GW production \cite{Addazi:2019mjh, Addazi:2016ksu} and memory effects \cite{Strominger:2014pwa, Pasterski:2015tva, Strominger:2017zoo, Addazi:2020obs, Aldi:2020qfu, Aldi:2021zhh}. 

Identifying (discrete) symmetries even in very special cases, such as eSTURBHs, can prove very useful in order to make further progress in this endeavour as well as in the scattering off D-branes, their bound-states \cite{DAppollonio:2010krb, Bianchi:2011se, DAppollonio:2015fly} or highly excited (coherent) string states \cite{Bianchi:2019ywd, Bianchi:2010dy, Gross:2021gsj} that can expose chaotic behaviour. Moreover the surprising connection with ${\cal N}=2$ SYM and, through the AGT correspondence, with 2-d Liouville CFTs may shed further light on the holographic correspondence in this contexts \cite{Kulaxizi:2018dxo, Kulaxizi:2019tkd, Parnachev:2020zbr}. 
  
\section*{Acknowledgements}
We acknowledge fruitful scientific exchange with  G.~Bonelli, G.~Bossard, D.~Consoli, M.~Cvetic, M.~Firrotta, F.~Fucito, A.~Grillo, F.~Morales, J.~Russo, R.~Savelli and A.~Tanzini. We would like to thank for the kind hospitality. We thank the MIUR PRIN contract 2020KR4KN2 "String Theory as a bridge between Gauge Theories and Quantum Gravity" and the INFN project ST\&FI "String Theory and Fundamental Interactions" for partial support.

\appendix

\section{Critical parameters for asymptotically flat KN BHs}\label{appA}

The critical conditions read:
\begin{align}\label{sistcrit}
&R(r_c)=(r_c^2-a\z_c)^2-b_c^2\Delta_r(r_c)=0\\\nn
&R'(r)=4 r_c(r_c^2-a\z_c)-b_c^2\Delta'_r(r_c)=0
\end{align}
must be supplemented with the condition 
\be\label{thetacrit}
b^2\sin^2\q-\left(\z+a\cos^2\q\right)^2 \ge 0
\ee
The solutions of \eqref{sistcrit} are:
\be
\z_c={r_c^2\over a}-{4 r_c\Delta_r(r_c)\over a \Delta'_r(r_c)}\quad,\quad b^2_c={16 r_c^2\Delta_r(r_c)\over[\Delta_r'(r_c)]^2}
\ee
Setting $X= \cos 2\theta$, $b_c/a=\beta$, $\zeta_c/a = \gamma$ and using
\be
 \sin^2\q = {1\over 2} (1-X) \quad , \quad \cos^2\q = {1\over 2} (1+X)
\ee
the condition  \eqref{thetacrit} become:
\be
2\beta^2(1-X) \ge (2\gamma + 1 + X)^2 
\ee
or
\be\label{cos}
X^2 + 2X(\beta^2+2\gamma + 1) + (2\gamma + 1)^2 - 2\beta^2 \le 0.
\ee
Since 
\be\label{discr}
\Delta = \beta^2 (\beta^2+4\gamma + 4)\ge0 
\ee 
then, setting $x=r/M$, $q=Q/M$ and $\a=a/M$, we obtain the following constraints:
\begin{align}
&\beta^2\ge0\Longrightarrow x\ge x_H^+=1+\sqrt{1-\a^2-q^2}\quad,\quad x\le x_H^-=1-\sqrt{1-\a^2-q^2}\\\nn
&\beta^2+4\gamma + 4\ge0 \Longrightarrow 2x^3+(-q^2-3)x^2+2q^2 x+\a^2\equiv f(x)\ge0
\end{align}
For consistency we must require (Schwarzschild) $0<\a^2+q^2<1$ (extremal KN).

\subsection{Non-extremal case: $M> \sqrt{a^2+Q^2}$}

The extremal points of $f(x)$ are:
\be
f'(x_e)=6x_e^2+2(-q^2-3)x_e+2q^2=0\Longrightarrow x_e^+=1\quad,\quad x_e^-={q^2\over3}.
\ee
It is very easy to see that $x_e^+$ is a minimum and $x_e^-$ is a maximum. 
%\begin{figure}[h!]
%\centering
%\includegraphics[width=.49\textwidth]{f(x)}
%\caption{The function f(x) for $\a=3/4$ and $q=1/5$. The zeros are in $x_1\simeq -0.375$, $x_2\simeq0.564$ and $x_3\simeq1.331$. The horizons are in $x_H^+\simeq1.630$ and in $x_H^-\simeq0.370$}\label{fig3}
%\end{figure}
It is crucial to note that in the allowed range of the parameters, $f(x)$ has always one negative and two positive real roots. Furthermore, if we denote the roots with $x_1$, $x_2$ and $x_3$ with $x_1<0<x_2<x_3$, the outer horizon $x_H^+$ is always larger the the biggest root $x_3$ and the inner horizon $x_H^-$ is always smaller than $x_2$.
%\begin{figure}[h!]
%\centering
%\includegraphics[width=.49\textwidth]{rh+}\hfil
%\includegraphics[width=.49\textwidth]{rh-}
%\caption{Respectively from the left to the right we show the 3D plot of $x_H^+$ compared with $x_3$ and of $x_H^-$ compared with $x_2$}\label{fig4}
%\end{figure}
So the positivity range of the discriminant \eqref{discr} is:
\be\label{photreg}
0\le x\le x_H^-\quad,\quad x_2\le x\le x_3\quad,\quad x\ge x_H^+
\ee
The solutions of \eqref{cos} are:
\be
X_{\pm} = - (\beta^2+2\gamma + 1) \pm \sqrt{\beta^2 (\beta^2+4\gamma + 4)}
\ee
are real with $X_-<X_+$ so that the allowed range is $[X_-,X_+]$ but one has to make sure that it intersects with $[-1,+1]$ since $X=\cos 2\theta$, after all.
%It is easy to check that $X_-<-1(<1)$, while $X_+<1$ is always true.
The condition $X_+\le 1$ is equivalent to 
\be 
\beta^2\ge \gamma^2 
\ee
Setting $m^2=\a^2+q^2$, $\beta$ is defined for $x>x_H^+$ and for $x<x_H^-$. The derivative of $\b$ is:
\be
\beta'={2(x^3-3x^2+3x-m^2)\over |\a|(x-1)^2\sqrt{\Delta_r}}
\ee
In the range $0<m^2<1$, $\beta$ has only one extremal point in
\be
x_e=1-\sqrt[3]{1-m^2}
\ee
which is a minimum and is always located between zero and $x_H^-$. Now we focus on:
\be
|\gamma|=\left|-{x(x^2-3x+2m^2)\over \a^2(x-1)}\right|.
\ee
The zeros of this function are:
\be
x_0^\pm={3\over 2}\pm\sqrt{{9\over4}-2m^2}.
\ee
It is very easy to show that 
\be
x_0^+>x_H^+\quad,\quad x_0^-<x_H^-\quad,\quad \text{for}\quad0<m^2<1
\ee
%\begin{figure}
%\centering
%\includegraphics[width=1\textwidth]{b&z}
%\caption{The functions $\b$ and $|\gamma|$ for $\a=0.3$ and $q=0.7$ (continuous lines) and the same functions in the extremal case (in particular we consider $\a=0.3$ and consequently $q=\sqrt{91}/10$). The region where $\b>|\gamma|$ is the photon region whose boundaries are $x_c^\pm$. In this example the horizons are located in $x_H^+\simeq1.648$ and in $x_H^-\simeq0.352$. The photon region is $2.197\le x_c \le2.989$.}\label{fig8}
%\end{figure}
In the non extremal case is always possible to find a photon region outside the outer horizon.

\subsection{Extremal case: $M=\sqrt{a^2+Q^2}$} 

The situation changes drastically in the extremal case in which $m^2=1$. The functions $\b$ and $\gamma$ reduce to:
\be
\b={2 x\over |\a|}\quad,\quad |\gamma|={|x(2-x)|\over \a^2}
\ee
%\begin{figure}
%\centering
%\includegraphics[width=1\textwidth]{b&ze}
%\caption{The functions $\b$ and $|\gamma|$ In the extremal case and for$\a=1/3$. The region where $\b>|\gamma|$ is the photon region whose boundaries are $x_c^\pm$. The photon region is ${4\over 3}\le x_c \le{8\over 3}.$}\label{fig8}
%\end{figure}
\be
\begin{cases}2x|\a|=x(2-x), & \mbox{if }0\le x\le 2\Longrightarrow x=0,\quad x=x_c^-=2-2|\a|  \\ 2x|\a|=x(x-2), & \mbox{if }x\ge2,\quad x\le0\Longrightarrow x=0,\quad x=x_c^+=2+2|\a|, 
\end{cases}
\ee
where obviously $x_c^-<x_c^+$. However, for consistency, we must require that the lower bound of the photon region is greater than the horizon:
\be
2-2|\a|>1\Longrightarrow |\a|={|a|\over \sqrt{a^2+Q^2}}<{1\over2}\Longrightarrow |a|\le{Q\over\sqrt{3}}
\ee

\section{Appendix: Translation Dictionary eSTURBH to eKN for $Q_1=Q_2=Q_3=Q_4=Q$ }\label{appB}

In this appendix we discuss the map between the parameters in the eKN metric and the ones in the eSTURBH. 
Denoting by $M$, $Q$, $a$, $r$, $\rho$ the parameters and variables in KN description and by $m$, $q$, $\alpha$, $u$, $\zeta$ the `corresponding' parameters and variables in CPS-STU description and setting $\sigma=\sinh\delta$ one finds (for $Q_1=Q_2=Q_3=Q_4=Q$)
\be
M= m(1+2\sigma^2) \quad Q=q=2m\sigma\sqrt{1+\sigma^2} \quad a=\alpha \quad m^2=M^2-Q^2
\ee
and
\be 
r = u +2m\sigma^2 \quad \Delta = r^2-2Mr+a^2+Q^2 = u^2-2mu+a^2 = \Delta_{CPS} \ee
\be
\rho^2 = r^2 + a^2\cos^2\theta = W =  [u +2m\sigma^2]^2+ a^2\cos^2\theta
\ee 
 \be
 \zeta^2 -2mu = u^2 + a^2\cos^2\theta -2mu = \Delta-a^2 + a^2\cos^2\theta = \Delta - a^2\sin^2\theta
 \ee
 Then
 \be
u{-}u_H = u{-}[m+\sqrt{m^2{-}a^2}] = r{-}m(1{+}2\sigma^2){-}\sqrt{M^2{-}Q^2{-}a^2}= r{-}M{-}\sqrt{M^2{-}Q^2{-}a^2} = r{-}r_H
\ee
that is 
\be
r = u -u_H+r_H 
\ee
In the extremal limit $M^2=a^2+Q^2$ one has $m^2=a^2$ with $u_H=m=a$ and $r_H=M=\sqrt{a^2+Q^2}$

\section{Heun Equation vs quantum SW curve}\label{appc}

In order to find the dictionary between HE for scalar fluctuations with $\mu^2L^2=-2$ around AdS KN in $D=4$ and quantum SW curve for ${\cal N}=2$ SYM theory with $G=SU(2)$ and $N_f=4=(2_L,2_R)$ we have to introduce the coordinate $y=-z$ and rewrite the second term in \eqref{he2} as follows:
\be\label{QH}
Q_H(y)={1-\delta^2\over 4y^2}+{2\delta-\delta^2\over 4(1+y)^2}+{2\delta-\delta^2\over 4(t+y)^2}+\frac{\delta ^2-\delta +2 p+\delta ^2 t-\delta  t+y(\delta ^2 -2 \delta)}{2 t y (y+1) (1+{1\over t}y)}
\ee
From quantum SW curve, we have:
\be\label{QSW}
Q_{SW}(y)=\sum_{i=1}^3{\sigma_i\over(y-y_i)^2}+{\nu_1+q y(\sigma_4-\sigma_1-\sigma_2-\sigma_3)\over y(1+y)(1+q y)}
\ee 
where $\{y_i=0,-1,-1/q\}$ and
\begin{align}
&\sigma_1={1\over 4}-{(m_1-m_2)^2\over 4\hbar^2}\quad,\quad\sigma_2={1\over 4}-{(m_1+m_2)^2\over 4\hbar^2}\\\nn
&\sigma_3={1\over 4}-{(m_3+m_4)^2\over 4\hbar^2}\quad,\quad\sigma_4={1\over 4}-{(m_3-m_4)^2\over 4\hbar^2}\\\nn
4\hbar^2\nu_1=(q-1)(\hbar^2&+4u)+2(m_1^2+m_2^2)+2q\Biggl[2m_3 m_4+(m_1+m_2)(m_3+m_4)-\hbar\sum_i m_i\Biggl]
\end{align}
By comparing the various terms of \eqref{QH} and \eqref{QSW}, we find the following dictionary:
\begin{align}\label{He3}
&\delta^2={(m_1-m_2)^2\over\hbar^2}={(m_3-m_4)^2\over\hbar^2}\\\nn
&(\delta-1)^2={(m_1+m_2)^2\over \hbar^2}={(m_3+m_4)^2\over \hbar^2}\\\nn
&t={1\over q}
\end{align}
If we take the square root of the first two relations in \eqref{He3}, we have to introduce the signs $\varepsilon_i=\pm 1$ for $i=1,...,4$. Finally the last element of the dictionary is:
\be
u={\hbar^2\over 2(t-1)}\Big\{-2p+(1-\delta)\Big[1+\delta+\varepsilon_3\varepsilon_4(1-\delta)+\varepsilon_3+\varepsilon_4)\Big]\Big\}\ee

\end{document}